\documentclass[aps,prb,preprint,superscriptaddress,showpacs]{revtex4-1}
\usepackage{graphicx} 
\usepackage{textcomp} %\textmu etc
\usepackage{hyperref} %clickable links
\usepackage{ifthen} % provides \ifthenelse test  
\usepackage{xifthen} % provides \isempty test
\usepackage{color} %colored
\usepackage{soul} %highlights with }
\usepackage{amsmath} %for equation* and many more
\usepackage{lineno}
\usepackage[english]{babel}
\usepackage{ucs}
\usepackage[utf8x]{inputenc}
\usepackage{amssymb}
\usepackage{amsfonts}
\usepackage{bm}
\usepackage{bbm}

\newcommand{\abs}[1]{\left|#1\right|}
\newcommand{\mm}[1]{\mathrm{#1}}
\newcommand{\ui}{\mathrm{i}}

\newcommand{\up}{\mathrm{p}}
\newcommand{\ub}{\mathrm{b}}
\newcommand{\us}{\mathrm{s}}
\newcommand{\ud}{\mathrm{d}}

\newcommand{\fig}[2][]{%
\ifthenelse{\isempty{#1}}
{Fig.~\ref{#2}}% if no subfigure is given
{Fig.~\ref{#2}(#1)}% else
}
\newcommand{\Fig}[2][]{%
\ifthenelse{\isempty{#1}}
{Figure~\ref{#2}}% if no subfigure is given
{Figure~\ref{#2}(#1)}% else
}

\begin{document}

%\preprint{}

\title{Finite time St\"uckelberg interferometry with nanomechanical modes}



\author{Maximilian\,J. Seitner}
\email{maximilian.seitner@uni-konstanz.de}
\affiliation{Departement of Physics, University of Konstanz, 78457 Konstanz, Germany}
\affiliation{Center for NanoScience (CeNS) and Fakult\"at f\"ur Physik, Ludwig-Maximilians-Universit\"at, Geschwister-Scholl-Platz 1,
M\"unchen 80539, Germany}
\author{Hugo Ribeiro}
\affiliation{Department of Physics, McGill University, Montreal, Quebec, H3A 2T8, Canada}
\author{Johannes K\"olbl}
\affiliation{Departement of Physics, University of Konstanz, 78457 Konstanz, Germany}
\author{Thomas Faust}
\affiliation{Center for NanoScience (CeNS) and Fakult\"at f\"ur Physik, Ludwig-Maximilians-Universit\"at, Geschwister-Scholl-Platz 1,
M\"unchen 80539, Germany}
\author{Eva M. Weig}
\affiliation{Departement of Physics, University of Konstanz, 78457 Konstanz, Germany}
\affiliation{Center for NanoScience (CeNS) and Fakult\"at f\"ur Physik, Ludwig-Maximilians-Universit\"at, Geschwister-Scholl-Platz 1,
M\"unchen 80539, Germany}

%\date{\today}

\begin{abstract}
St\"uckelberg interferometry describes the interference of two strongly coupled modes during a double passage through an avoided energy level crossing. In this work, we experimentally investigate finite time effects in St\"uckelberg interference and provide an exact analytical solution of the St\"uckelberg problem. Approximating this solution in distinct limits reveals uncharted parameter regimes of St\"uckelberg interferometry. Experimentally, we study these regimes using a purely classical, strongly coupled nanomechanical two-mode system of high quality factor. The classical two-mode system consists of the in-plane and out-of-plane fundamental flexural mode of a high stress silicon nitride string resonator, coupled via electric gradient fields. The dielectric control and microwave cavity enhanced universal transduction of the nanoelectromechanical system allows for the experimental access to all theoretically predicted St\"uckelberg parameter regimes. We exploit our experimental and theoretical findings by studying the onset of St\"uckelberg interference in dependence of the characteristic system control parameters and obtain characteristic excitation oscillations between the two modes even without the explicit need of traversing the avoided crossing. The presented theory is not limited to classical mechanical two-mode systems but can be applied to every strongly coupled (quantum) two-level system, for example a spin-1/2 system or superconducting qubit.
\end{abstract}

%\pacs{???}
\maketitle

\section{Introduction}
\label{intro}
Strongly coupled nanomechanical resonators have proven themselves as prominent testbed for the
investigation of various fundamental physical concepts. The recent studies of, for example,
non-classical correlations\cite{2016_Riedinger_Nature,2016_Sudhir_arxiv}, quantum
back-action\cite{2016_Spethmann_NatPhys}, quantum squeezing\cite{2016_Wollman_Science} and
topological effects\cite{2016_Xu_arxiv} in different nanomechanical systems demonstrate in an
outstanding way the scientific impact of hybrid-mechanical systems. In addition, the high level of
control over such coupled resonators allows for the realization of ultrasensitive vectorial force
sensors\cite{2016_Rossi_arxiv,2016_Mercier_arxiv} and $\Lambda$-type three level
systems\cite{2016_Okamoto_APL}.

Recently, this high level of control led to the demonstration of classical St\"uckelberg
interference of two strongly coupled nanomechanical resonator modes\cite{2016_Seitner_NatPhys}. This
coherent transfer of energy has originally been studied in a broad range of quantum systems
including, e.g., spin-1/2 systems\cite{2010_Petta_Science,Gaudreau2012,2013_Ribeiro_PRL} and
superconducting
qubits\cite{2005_Oliver_Science,2006_Silanpaa_PRL,2009_LaHaye_Nature,2010_Shevchenko_Review,2012_Shevchenko_PRB},
amongst many others. Typically, the coherent dynamics of a two-level system in the configuration
proposed by St\"uckelberg\cite{1932_Stuckelberg} is theoretically modeled by an infinite time
approach, the so-called adiabatic impulse model\cite{2010_Shevchenko_Review}. Following this model,
the interference of two quantum states during a double passage through an avoided level crossing
solely relies on the mutual coupling and is independent of the exact time evolution of the two
states in the vicinity of the avoided crossing. In this work, we go well beyond this simple
approximation and show that the adiabatic impulse model represent just one particular limit, the
infinite time limit, of the full St\"uckelberg problem\cite{1932_Stuckelberg}. We provide an exact
analytical solution to the problem which captures the importance of finite time effects. By means of
asymptotic approximations of the exact finite time solution, we identify up to six different
parameter regimes of St\"uckelberg interferometry. Experimentally, we demonstrate that a classical
strongly coupled nanomechanical two-mode system\cite{2013_Faust_NatPhys,2016_Seitner_NatPhys} allows
for the investigation of all discussed asymptotic regimes due to high mechanical quality factors and
hence lifetimes of the coherent mechanical modes in the millisecond
regime\cite{2013_Faust_NatPhys,2016_Seitner_NatPhys}.

The manuscript is organized as follows. Following this introduction (\ref{intro}), the
nanoelectromechancial system as well as the experimental techniques are introduced in the second
part (\ref{system}). In sections \ref{coupling} and \ref{cqt}, we derive an exact analytical
solution of the St\"uckelberg problem, taking advantage of the conformity of
classical and quantum interference in this particular problem\cite{2016_Seitner_NatPhys}.
Additionally, the asymptotic limits of the exact solution are derived (appendices\,\ref{app:asymptotic}\,\&\,\ref{sec:returnPloc}) which
allows for a quantification of characteristic parameter regimes in St\"uckelberg interferometry. In chapter\,\ref{aqt}, we explicitely derive the asymptotic long time limit of the analytical solution.
Chapter \ref{st} provides a brief summary of previous, approximative theoretical approaches of
St\"uckelberg interferometry and establishes the link to the presented exact analytical solution.
Section \ref{exp} compares the different theories to the experimentally observed classical
St\"uckelberg oscillations of a strongly coupled nanomechanical system. In the last part
(\ref{sec:conc}), we summarize the results.

\section{The nanoelectromechanical system}
\label{system}
\subsection{Experimental set-up}
\begin{figure}[!htb]
\includegraphics{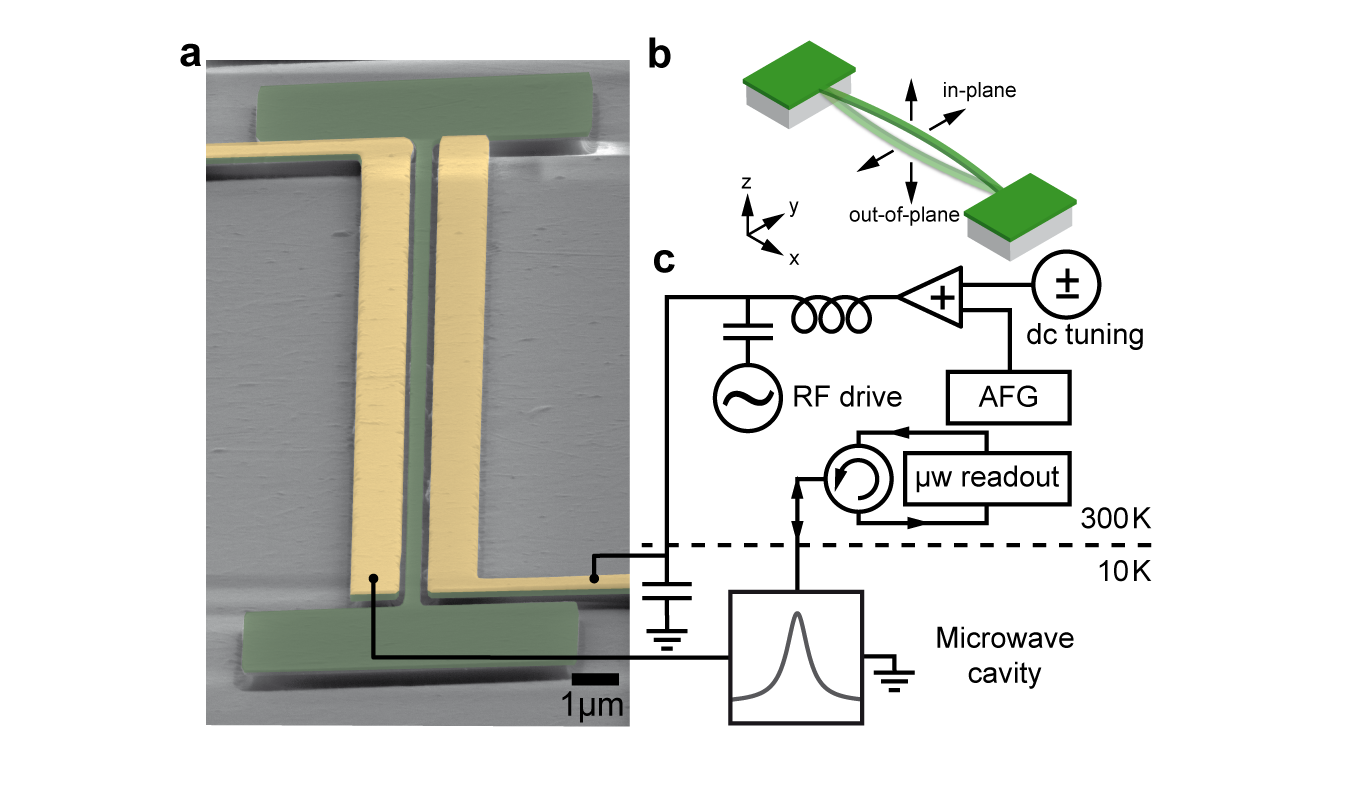}
\caption{\label{Fig_1} Schematic experimental set-up. \textbf{a},\,False color scanning electron micrograph of a 50\,\textmu m long, 270\,nm wide and 100\,nm thick silicon nitride string resonator (green) in oblique view. The mechanical resonator is flanked by two 1\,\textmu m wide gold electrodes (yellow), which are processed on top of the silicon nitride and form a capacitor providing dielectric drive, tuning and detection as well as mode coupling. \textbf{b},\,Schematic illustration of the two orthogonally polarized fundamental flexural vibration modes of the silicon nitride string resonator. The oscillation in z-direction, perpendicular to the sample plane, is referred to as out-of-plane oscillation, whereas the oscillation in y-direction, parallel to the sample plane, is referred to as in-plane oscillation. \textbf{c},\, Schematic equivalent circuit diagram of the electrical drive, tuning and heterodyne detection scheme. The voltage ramp is added to the DC tuning voltage by a summation amplifier and combined with the resonant sinusoidal RF drive tone at a bias tee. The combined voltages are applied to one of the gold electrodes versus the ground of the microwave cavity. The bypass capacitor acts as a ground for the microwave cavity. The microwave cavity is driven on resonance and the signal is read-out via a heterodyne IQ-mixing technique, demodulating the sidebands induced by the oscillation of the nanomechanical resonator. }
\end{figure}

We study self-interference of a classical nanomechanical two-mode system using two samples of the same basic design. Sample\,A is investigated in a pulse-tube cryostat at a temperature of 10\,K which serves solely for temperature stabilization. The experiments on sample\,B are conducted at room temperature. Independent of the ambient temperature, both samples operate deeply in the classical regime and do not exhibit quantum mechanical properties\cite{2013_Faust_NatPhys,2016_Seitner_NatPhys}. The samples consist of freely-suspended and doubly clamped silicon nitride (SiN) string resonators, fabricated in a top-down approach from a high-stress silicon nitride film on a fused silica substrate. The 100\,nm thick and 270\,nm wide silicon nitride strings exhibit a high tensile pre-stress of 1.46\,GPa resulting from the LPCVD deposition process of the SiN atop the fused silica wafer. The high tensile pre-stress translates into high mechanical quality factors up to $Q\approx 500,000$ at mechanical resonance frequencies of $\omega_\mathrm{m}/2\pi\approx 6.5$\,MHz at room temperature. Sample\,A consists of a 50\,\textmu m long string resonator, whereas on sample\,B we study a 55\,\textmu m long string. As depicted in Fig.\,\ref{Fig_1}\,a and Fig.\,\ref{Fig_1}\,b, the string resonators exhibit two fundamental flexural vibration modes with orthogonal mode polarizations, namely one perpendicular to the sample plane (out-of-plane) and one parallel to the sample plane (in-plane). For dielectric control and transduction of the string resonators (cf. Fig.\,\ref{Fig_1}\,c) we process two gold electrodes adjacent to the SiN strings, which form a capacitor and are connected to a microwave cavity\cite{2012_Faust_NatComm} via a bond wire. The oscillation of the dielectric silicon nitride string between the gold electrodes periodically modulates the capacitance. This change in capacitance in turn modulates the $\lambda /4$ microstrip cavity signal with resonance frequency at approximately $\Omega_\mathrm{c}/2\pi\approx3.6$\,GHz by producing sidebands on the cavity signal at $\Omega_{\pm}=\Omega_\mathrm{c}\pm \omega_\mathrm{m}$, where $\omega_\mathrm{m}/2\pi\approx 6.5$\,MHz denotes the mechanical resonance frequency. The modulation induced sidebands are not resolved but can be demodulated via a heterodyne in-phase-quadrature mixing technique\cite{2012_Faust_NatComm} before subsequent low-pass filtering and amplification. Finally, the demodulated signal is captured using a spectrum analyzer.
In addition to the described microwave cavity enhanced heterodyne dielectric detection, the gold electrodes are used at the same time for dielectric actuation and control of the mechanical resonance\cite{2012_Rieger_APL}. Applying a DC bias to one of the electrodes induces an electric polarization in the dielectric silicon nitride string, which, in turn, couples to the gradient of the inhomogeneous electric field, generating a gradient force. Adding a resonant sinusoidal RF drive tone with frequency $\omega_\mathrm{m}/2\pi$ to the DC voltage at a bias tee results in a periodic force which drives the vibrational resonance of the nanomechanical silicon nitride string resonator\cite{2009_Quirin_Nature}. Approximating the induced electrical polarization by a dipole moment\cite{2009_Quirin_Nature,2012_Rieger_APL}, its magnitude scales linearly with the applied DC voltage. Since the electric field gradient is also directly proportional to the DC voltage, the resonance frequency of the nanomechanical string resonator shifts quadratically with the applied DC bias\cite{2012_Rieger_APL}. By means of careful sample design, the in-plane polarized vibration mode can be engineered to shift downwards in resonance frequency with increasing DC bias, whereas the out-of-plane polarized resonance tunes towards higher resonance frequencies\cite{2012_Rieger_APL}. Thereby, the inherent resonant frequency off-set between the two orthogonally polarized vibration modes, which arises from the rectangular cross-section of the nanomechanical string, can be compensated. Near resonance, the two modes hybridize into normal modes\cite{2016_Rossi_arxiv,2016_Mercier_arxiv,2012_Faust_PRL} of the strongly coupled system, diagonally polarized along $\pm 45$\textdegree with respect to the sample plane. The strong coupling, mediated by the inhomogeneous electric field\cite{2012_Faust_PRL,2013_Faust_NatPhys}, is reflected by the pronounced avoided crossing of the two mechanical modes with level splitting $\Delta/2\pi$ as depicted in Fig.\,\ref{Fig_2}\,a.
\subsection{Measurement scheme}
\label{meas}
\begin{figure}[!htb]
\includegraphics{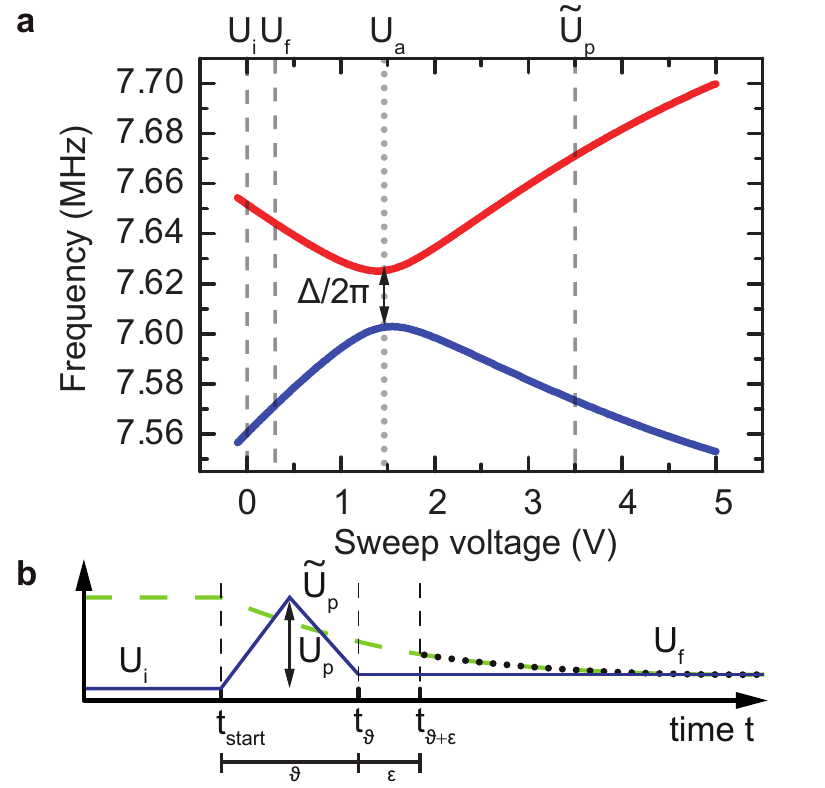}
\caption{\label{Fig_2} Avoided crossing of sample A and voltage ramp sequence. \textbf{a},\,Avoided energy level crossing of the two frequency branches (mode\,1: blue, mode\,2: red) which stem from the two orthogonally polarized flexural modes. Gray dashed lines indicate the initialization voltage $U_\mathrm{i}$, the read-out voltage $U_\mathrm{f}$ and an exemplary absolute peak voltage $\widetilde{U}_\mathrm{p}=U_\mathrm{i}+U_\mathrm{p}$. The gray dotted line represents the avoided crossing voltage $U_\mathrm{a}$, where the two modes exhibit a frequency splitting of $\Delta/2\pi$. \textbf{b},\,Temporal evolution of the voltage ramps (blue solid line) defined by the sweep voltage. The ramp starts at $t=t_\mathrm{start}$. The sweep voltage is increased from zero to peak voltage $U_\mathrm{p}$ at voltage sweep rate $\beta$, which increases the absolute voltage from $U_\mathrm{i}$ to $\widetilde{U}_\mathrm{p}=U_\mathrm{i}+U_\mathrm{p}$. At the apex of the triangular voltage ramp (peak voltage $U_\mathrm{p}$), the sweep voltage is decreased at the same rate to the read-out voltage $U_\mathrm{f}$, which is approached at time $t=t_\vartheta$. Hence, the complete triangular voltage ramp has a duration of $\vartheta$. Note that the read-out voltage $U_\mathrm{f}$ is off-set from the initialization voltage $U_\mathrm{i}$ as explained in the text. As a consequence, the sweep voltage does not return to zero. The ring-down of the mechanical signal power (green dashed line) is measured after a delay $\varepsilon$ (at time $t=t_\mathrm{\vartheta+\epsilon}$), and a fit (black dotted line) is used to extract its magnitude at time $t=t_\vartheta$. The measured return signal is normalized to the mechanical signal power at $t=t_\mathrm{start}$.}
\end{figure}

In this work, we study the effects of finite times in classical St\"uckelberg interferometry. In general, St\"uckelberg interference\cite{1932_Stuckelberg} occurs during a double passage through an avoided energy level crossing within the coherence time of the strongly coupled system. Both energy branches accumulate phase during the double passage, giving rise to self-interference. This brings about interference fringes depending on the difference in the accumulated phase. The probability to find the system either in the upper or the lower energy branch after the double passage oscillates in dependence of the level splitting, the traversal time as well as the initialization and turning point\cite{1932_Stuckelberg,2010_Shevchenko_Review}.\\
Experimentally, we realize the double passage of the avoided crossing using fast triangular voltage ramps\cite{2016_Seitner_NatPhys}. The voltage ramps are provided by an arbitrary function generator (AFG) and combined with the fixed DC tuning voltage at a summation amplifier. A detailed description of the ramps can be found in appendix\,\ref{ramp}. In the following, we focus solely on the measurement principle which is equivalent for the investigation of sample\,A at $10$\,K and sample\,B at room temperature. The detailed experimental parameters of the respective samples are discussed in chapter\,\ref{exp}. Note that the presented voltage ramp sequence\cite{2016_Seitner_NatPhys} is analogous to the one employed by Sun \textit{et al.}\cite{2011_Sun_PRB} and differs from the frequently performed periodic driving schemes in St\"uckelberg interferometry experiments\cite{2010_Shevchenko_Review}. The schematic sequence of the applied voltage ramp is depicted in Fig.\,\ref{Fig_2}\,b. The system is initialized in the lower frequency branch at $\omega_1(U_\mathrm{i})/2\pi$ by the application of a resonant sinusoidal RF drive tone. Hereby, $U_\mathrm{i}$ denotes the initialization voltage to which the DC tuning voltage is set during a St\"uckelberg experiment. Note that this voltage corresponds to a sweep voltage of zero. The sweep voltage defines the additional ramp voltage provided by the AFG. At $t=t_\mathrm{start}$, the fast voltage ramp is turned on and detunes the system from the resonant drive at $\omega_1(U_\mathrm{i})/2\pi$. From this time on, the mechanical resonator is not driven any more and its oscillation decays exponentially (green dashed line in Fig.\,\ref{Fig_2}\,b). Note that the mechanical energy decays on a larger timescale than the duration of the fast voltage ramp. The sweep voltage ramps the system from $U_\mathrm{i}$ through the avoided crossing at voltage $U_\mathrm{a}$, up to the absolute peak voltage $\widetilde{U}_\mathrm{p}=U_\mathrm{i}+U_\mathrm{p}$ and back to the read-out voltage $U_\mathrm{f}$ during time $\vartheta$. At time $t=t_{\vartheta+\varepsilon}$, we start to measure the exponential decay of the mechanical oscillation at frequency $\omega_1(U_\mathrm{f})/2\pi$ in the lower branch at the read-out voltage $U_\mathrm{f}$. The return signal needs to be measured at $U_\mathrm{f}$ since the drive at  $\omega_1(U_\mathrm{i})/2\pi$ cannot be turned off during the experiment. Hence, a measurement at $U_\mathrm{i}$ would lead to another excitation of the mode and therefore destroy the interference. Additionally, the exponential decay of the return signal power needs to be measured with a temporal off-set $\varepsilon$ to avoid transient effects. The exponential decay is extrapolated back to the time $t_\vartheta$ where the voltage ramp ended via a fit and the resulting return signal power is normalized to the signal power at the time of initialization of the resonance ($t=t_\mathrm{start}$). This normalization process can lead to return probabilities exceeding a value of unity due to experimental scatter and different characteristic signal power heights at the initialization and read-out voltage. Consequently, we use the term \textit{normalized squared return amplitude} for the experimental data instead of return probability.\\
For each particular measurement, the voltage ramp has a fixed voltage sweep rate $\beta$ and fixed peak voltage $U_\mathrm{p}$. The experiment is repeated for a set of different voltage sweep rates at a fixed peak voltage. Subsequently, the peak voltage is changed and the measurement procedure is repeated. In this way, we investigate classical St\"uckelberg interferometry as a function of sweep speed and sweep distance which can be absorbed into a single variable, namely time.\\
In previous approaches\cite{2010_Shevchenko_Review}, St\"uckelberg interferometry has been investigated in the limit of infinite times. That means the initialization and turning point on the left and the right hand side of the avoided crossing are far away from the point of maximum coupling, which is at voltage $U_\mathrm{a}$ where the level splitting is $\Delta/2\pi$, compared to the characteristic time-scale of the system. This infinite time approximation is referred to as the  \textit{adiabatic impulse model}\cite{2010_Shevchenko_Review} and is summarized in chapter\,\ref{st}. In this work, we go beyond this approximation via the investigation of finite time effects. Experimentally, we interface this regime by turning points, i.e. peak voltages, close to or even before the avoided crossing, which still results in characteristic St\"uckelberg oscillations.
\section{Finite-time theory}
\label{theo}
\subsection{Theory of strongly coupled modes}
\label{coupling}
%}
In order to theoretically model the two modes in the strong coupling regime, we follow the work of
Novotny \textit{et al.}\cite{2010_Novotny_AJP} and write the system as two coupled differential
equations:
\begin{equation}
	\begin{aligned}
		m\ddot{u}_1 (t) + m \omega_1^2 u_1 (t) + \kappa  \left[ u_1 (t) - u_2 (t)\right] &=
		0 \\
		m\ddot{u}_2 (t) + m \omega_2^2 u_2 (t) + \kappa \left[ u_2 (t) - u_1 (t) \right] & = 0
	\end{aligned}
\label{eq:couple}
\end{equation}
where $m=m_0/2$ denotes the effective mass of the resonator with physical mass $m_0$, $u_j$ ($j=1,\,2$) the displacement of
mode $j$, $\omega_j=\sqrt{k_j/m}$ the respective angular resonance frequency, $k_j$ the spring constant of mode $j$, and $\kappa$ the coupling constant
between the two modes. Using the ansatz $u_j (t) = u_{0,j}\exp(-i\omega_{\pm} t)$ in Eq.~\eqref{eq:couple} yields the resonance frequencies of the two normal modes in the coupled
system:
\begin{equation}
\label{freqs}
\omega_{\pm}=\frac{\omega_\mathrm{1}^2+\omega_\mathrm{2}^2\pm\sqrt{(\omega_\mathrm{1}^2-\omega_\mathrm{2}^2)^2+4\Delta^2\omega_\mathrm{1}\omega_\mathrm{2}}}{2}
\end{equation}
Here, we define the level splitting
\begin{equation}
\label{split}
\Delta=\left|\lambda\right|=\frac{\kappa}{2m\sqrt{\omega_\mathrm{1}\omega_\mathrm{2}}}=\omega_+-\omega_-,
\end{equation}
where the coupling $\lambda$, in general, can be complex valued. If the level splitting exceeds the dissipation in the
system, namely the linewidth of the mechanical resonances, the modes can coherently exchange energy
on a faster timescale than the energy decay. This strong coupling regime allows
for the investigation of time dependent phenomena, like non-adiabatic Landau-Zener
tunneling\cite{2010_Novotny_AJP,2012_Faust_PRL} in the classical regime, coherent dynamics of classical two-mode
systems\cite{2013_Okamoto_NatPhys,2013_Faust_NatPhys,2014_Shkarin_PRL} and classical state
interferometry\cite{2016_Seitner_NatPhys}.
\subsection{Finite-time St\"uckelberg interferometry}
\label{cqt}

We look for a solution of Eq.~\eqref{eq:couple} in the experimentally relevant limit where
$\kappa/k_j \ll 1$, $j \in \{1,2\}$. This suggests to look for solutions of the form $u_j (t) = c_j
(t) \exp[i \tilde{\omega}_1 t]$ with $c_j (t)$ a normalized amplitude, i.e. $\abs{c_1 (t)}^2 +
\abs{c_2 (t)}^2 =1$, and we define $\tilde{\omega}_j = \sqrt{(k_j + \kappa)/m}$. By replacing our
ansatz for $u_j (t)$ in Eq.~\eqref{eq:couple}, we find 
\begin{equation}
	\begin{aligned}
		\ddot{c}_1 (t) + 2 i \tilde{\omega}_1 \dot{c}_1 (t) - \frac{\kappa}{m} c_2 (t) &=0 \\
		\ddot{c}_2 (t) + 2 i \tilde{\omega}_1 \dot{c}_2 (t) + (\tilde{\omega}_2^2 -
		\tilde{\omega}_1^2)c_2 (t)- \frac{\kappa}{m} c_1 (t) &=0,
	\end{aligned}
	\label{eq:fulldiffeqc1c2}
\end{equation}
Since the amplitudes $c_j (t)$ are slowly varying in time compared to the oscillatory function
$\exp[i \tilde{\omega}_1 t]$ (see for instance Ref.~\onlinecite{2009_Shore_AJP}), one can neglect
the second derivatives $\ddot{c}_j (t)$ in Eq.~\eqref{eq:fulldiffeqc1c2}. Thus, the evolution of the
normalized amplitudes is described by 
\begin{equation}
	i \dot{\mathbf{c}}(t) = H (t) \mathbf{c} (t),
	\label{eq:c1c2_matrix}
\end{equation}
where we have defined $\mathbf{c} (t) = (c_1 (t)\,c_2(t))^{\mm{T}}$ and
\begin{equation}
	H(t) =
	\begin{pmatrix}
		0 && \frac{\lambda}{2}\\
		\frac{\lambda}{2} && -\alpha t 
	\end{pmatrix},
	\label{eq:dynmat}
\end{equation}
with $\lambda = \kappa/(m \tilde{\omega}_1)$. To obtain Eq.~\eqref{eq:dynmat}, we have used that in
the vicinity of the avoided crossing $\tilde{\omega}_2\simeq \tilde{\omega}_1$. This yields
$(\tilde{\omega}_2^2-\tilde{\omega}_1^2)/(2 \tilde{\omega}_1) \simeq
\tilde{\omega}_2-\tilde{\omega}_1$ and we assume that the difference in frequency is changed in time
such that $\tilde{\omega}_2 - \tilde{\omega}_1 \simeq \alpha t$, where $\alpha$ denotes the frequency sweep rate. Note that we employ the frequency sweep rate $\alpha$ in the theory which is converted to the experimentally accessible voltage sweep rate $\beta$ using the conversion factor $\zeta$ from frequency to voltage (cf. section\,\ref{sec:st} and Ref.~\onlinecite{2016_Seitner_NatPhys}):
\begin{equation}
\alpha = 2\pi \times \zeta \times \beta .
\label{eq:conv}
\end{equation}

By applying the time-dependent unitary transformation
\begin{equation}
	S(t) =\exp\left[i\frac{\alpha}{4} t^2\right]\mathbbm{1}_2,
	\label{eq:Slz}
\end{equation}
to Eq.~\eqref{eq:c1c2_matrix}, with $\mathbbm{1}_2$ the two-dimensional identity operator, we find
that the transformed amplitudes obey the differential equation
\begin{equation}
	i \dot{\mathbf{c}}_{\mm{LZ}} (t) = H_{\mm{LZ}} (t)\mathbf{c}_{\mm{LZ}} (t) 
	\label{eq:c1c2_lzsm}
\end{equation}
where the dynamical matrix 
\begin{equation}
	H_{\mm{LZ}} (t) = \frac{1}{2}
	\begin{pmatrix}
		\alpha t && \lambda\\
		\lambda && -\alpha t
	\end{pmatrix}
	\label{eq:c1c2_matrix_lzsm}
\end{equation}
is analog to the representation of the Landau-Zener Hamiltonian in the basis of diabatic states.
Thus, the finite-time solution of the Landau-Zener problem~\cite{1996_Vitanov_PRA} is also a
solution of Eq.~\eqref{eq:c1c2_lzsm}. By analogy with the quantum mechanical case, we express the
solution using a classical flow $\varphi_{\mm{LZ}}$ (see for instance Hamiltonian flow in
Ref.~\onlinecite{arnold1989}),
\begin{equation}
	\mathbf{c}_{\mm{LZ}} (t) = \varphi_{\mm{LZ}} (t,t_\ui) \mathbf{c}_{\mm{LZ}} (t_\ui),
	\label{eq:flowlzsm}
\end{equation}
with $\mathbf{c}_{\mm{LZ}} (t_\ui)$ the initial condition of the system and 
\begin{equation}
	\varphi_{\mm{LZ}} (t,t_\ui) =
	\begin{pmatrix}
		\varphi_{\mm{LZ},11} (t,t_\ui) && \varphi_{\mm{LZ},12} (t,t_\ui) \\
		-\varphi_{\mm{LZ},12}^\ast (t,t_\ui) && \varphi_{\mm{LZ},11}^\ast (t,t_\ui)
	\end{pmatrix}.
	\label{eq:flowlzsmmatrix}
\end{equation}
We have 
\begin{equation}
	%\begin{aligned}
	%	&\varphi_{\mm{LZ},11}(t, t_\ui) =\\
	%	&\frac{\Gamma\left(1+i\frac{\eta^2}{4} \right)}{\sqrt{2\pi}}
	%	\left[D_{-1-i\frac{ \eta^2}{4}}\left(e^{-i\frac{3\pi}{4}}\tau_\ui\right)
	%	D_{-i\frac{\eta^2}{4}}\left(e^{i\frac{\pi}{4}}\tau\right)\right.\\
	%	&\phantom{={}}+\left. D_{-1-i\frac{\eta^2}{4}}\left(e^{i\frac{\pi}{4}}\tau_\ui\right)
	%	D_{-i\frac{\eta^2}{4}}\left(e^{-i\frac{3\pi}{4}}\tau\right)\right]
	%\end{aligned}
	\varphi_{\mm{LZ},11}(t, t_\ui) = 
		\frac{\Gamma\left(1+i\frac{\eta^2}{4} \right)}{\sqrt{2\pi}}
		\left[D_{-1-i\frac{ \eta^2}{4}}\left(e^{-i\frac{3\pi}{4}}\tau_\ui\right)
		D_{-i\frac{\eta^2}{4}}\left(e^{i\frac{\pi}{4}}\tau\right) 
		+ D_{-1-i\frac{\eta^2}{4}}\left(e^{i\frac{\pi}{4}}\tau_\ui\right)
		D_{-i\frac{\eta^2}{4}}\left(e^{-i\frac{3\pi}{4}}\tau\right)\right]
	\label{eq:phi11}
\end{equation}
and
\begin{equation}
	%\begin{aligned}
	%	&\varphi_{\mm{LZ},12}(t, t_\ui)=\\
	%	&\frac{\Gamma\left(1+i\frac{\eta^2}{4} \right)}{\sqrt{2\pi}}\frac{2}{\eta}
	%	e^{-i\frac{\pi}{4}}
	%	\left[D_{-i\frac{ \eta^2}{4}}\left(e^{-i\frac{3\pi}{4}}\tau_\ui\right)
	%	D_{-i\frac{ \eta^2}{4}}\left(e^{i\frac{\pi}{4}}\tau\right)\right.\\
	%	&\phantom{={}}-\left.D_{-i\frac{\eta^2}{4}}\left(e^{i\frac{\pi}{4}}\tau_\ui\right)
	%	D_{-i\frac{ \eta^2}{4}}\left(e^{-i\frac{3\pi}{4}}\tau\right)\right],
	%\end{aligned}
	\varphi_{\mm{LZ},12}(t, t_\ui)=\\
	\frac{\Gamma\left(1+i\frac{\eta^2}{4} \right)}{\sqrt{2\pi}}\frac{2}{\eta}
	e^{-i\frac{\pi}{4}}
	\left[D_{-i\frac{ \eta^2}{4}}\left(e^{-i\frac{3\pi}{4}}\tau_\ui\right)
	D_{-i\frac{ \eta^2}{4}}\left(e^{i\frac{\pi}{4}}\tau\right)
	- D_{-i\frac{\eta^2}{4}}\left(e^{i\frac{\pi}{4}}\tau_\ui\right)
	D_{-i\frac{ \eta^2}{4}}\left(e^{-i\frac{3\pi}{4}}\tau\right)\right],
	\label{eq:phi12}
\end{equation}
where we have introduced dimensionless quantities by defining $\tau = \sqrt{\alpha} t$ and $\eta = \lambda/\sqrt{\alpha}$. Finally, the flow
describing the evolution of $c_1(t)$ and $c_2 (t)$ is given by 
\begin{equation}
	\varphi(t, t_\ui) = \exp\left[\frac{i}{4}\left(\tau^2 - \tau_\ui^2\right)\right]
	\varphi_{\mm{LZ}} (t, t_\ui).
	\label{eq:flowc1c2}
\end{equation}

The flow $\varphi(t, t_\ui)$ allows us to write in a simple way the state of the system after
multiple passages through the avoided crossing. In particular, for a double passage we have 
\begin{equation}
	\mathbf{c} (t) = \varphi_\ub (t, -t_\up) \varphi (t_\up, t_\ui) \mathbf{c}(t_\ui),
	\label{eq:solstueckelberg}
\end{equation}
with $\varphi_\ub (t, t_\ui) = \sigma_x \varphi (t, t_\ui) \sigma_x$ describing the evolution of the
system during the back sweep, $\sigma_x$ denotes the Pauli matrix in $x$-direction, and $t_\up$
labels the time at which the forward (backward) sweep stops (starts). The fact that $\varphi_\ub (t,
t_\ui) = \sigma_x \varphi (t, t_\ui) \sigma_x$ can be understood by noticing that during the back
sweep the frequency of mode $1$ ($2$) decreases (increases) while it increases (decreases) during
the forward sweep (see Fig.\,\ref{Fig_2}\,a).  

From
Eq.~\eqref{eq:solstueckelberg}, one obtains the return probability to mode 1,
\begin{equation}
	P_{1\to 1} = \abs{\varphi_{11} (t_\up,t_\ui) \varphi_{11}^\ast(t, -t_\up) +
	\varphi_{12}^\ast (t_\up,t_\ui) \varphi_{12}^\ast(t, -t_\up)}^2.
	\label{eq:returnprob}
\end{equation}

\subsection{Asymptotic solution for the long-time limit}
\label{aqt}
\subsubsection{Long time limit}
\label{sec:longtime}
In this section, we show how to obtain an approximate form of Eq.~\eqref{eq:returnprob} in the
long-time limit, i.e. $\tau, \abs{\tau_\ui}, \tau_\up \gg 1$. Using the respective asymptotic expansion of the
parabolic cylinder function (see appendix~\ref{app:asymptotic}), we find 
\begin{equation}
	\begin{aligned}
		\varphi_{\mm{LZ},11}(t, t_\ui) &\simeq \sqrt{1-e^{-\pi \frac{\eta^2}{2}}}
		\left[\sin[\theta(\abs{\tau_\ui})] \cos[\theta(\tau)]
		e^{i\left(\xi(\abs{\tau_\ui}) + \xi
		(\tau) + \frac{\pi}{4} - \mm{arg}\left[\Gamma\left(1 + i
		\frac{\eta^2}{4}\right)\right]\right)} 
		\right. \\
		&\phantom{= \sqrt{1-e^{-\pi \frac{\eta^2}{2}}} {}}
		\left .
		+ \cos[\theta(\abs{\tau_\ui})] \sin[\theta(\tau)] 
		e^{-i\left(\xi(\abs{\tau_\ui}) + \xi(\tau) 
		+ \frac{\pi}{4} - \mm{arg}\left[\Gamma\left(1 + i
		\frac{\eta^2}{4}\right)\right]\right)} \right] \\
		&\phantom{={}}
		+ e^{-\pi \frac{\eta^2}{4}} \left[\cos[\theta(\abs{\tau_\ui})] \cos[\theta(\tau)]
		e^{i[\xi(\abs{\tau_\ui}) - \xi(\tau)]} 
		- \sin[\theta(\abs{\tau_\ui})]
		\sin[\theta(\tau)] e^{-i[\xi(\abs{\tau_\ui}) - \xi(\tau)]}\right]
	\end{aligned}
	\label{eq:asymptphi11}
\end{equation}
and
\begin{equation}
	\begin{aligned}
		\varphi_{\mm{LZ},12}(t, t_\ui) &\simeq \sqrt{1-e^{-\pi \frac{\eta^2}{2}}}
		\left[-\sin[\theta(\abs{\tau_\ui})] \sin[\theta(\tau)]
		e^{i\left(\xi(\abs{\tau_\ui}) + \xi (\tau) + \frac{\pi}{4} - \mm{arg}\left[\Gamma\left(1 + i
		\frac{\eta^2}{4}\right)\right]\right)} \right. \\
		&\phantom{= \sqrt{1-e^{-\pi \frac{\eta^2}{2}}} {}}
		\left.
		+ \cos[\theta(\abs{\tau_\ui})]  \cos[\theta(\tau)]
		e^{-i\left(\xi(\abs{\tau_\ui}) + \xi(\tau) 
		+ \frac{\pi}{4} - \mm{arg}\left[\Gamma\left(1 + i
		\frac{\eta^2}{4}\right)\right]\right)} \right] \\
		&\phantom{={}}
		+ e^{-\pi \frac{\eta^2}{4}} \left[\cos[\theta(\abs{\tau_\ui})] \sin[\theta(\tau)]
		e^{-i[\xi(\abs{\tau_\ui}) - \xi(\tau)]} 
		- \sin[\theta(\abs{\tau_\ui})]
		\cos[\theta(\tau)] e^{i[\xi(\abs{\tau_\ui}) - \xi(\tau)]}\right].
	\end{aligned}
	\label{eq:asymptphi12}
\end{equation}
The functions $\cos[\theta(\tau)], \sin[\theta(\tau)]$, and $\xi(\tau)$ are defined in appendix\,\ref{app:asymptotic} and $\Gamma(z)$ is the gamma function.\\
Substituting these expressions in Eq.~\eqref{eq:returnprob} yields an expression for the return
probability that is valid in the long time limit. Note that the expansions in Eq.~\eqref{eq:asymptphi11} and Eq.~\eqref{eq:asymptphi12} are also valid for the softer criteria $\tau^2 + \eta^2/4 > 1$, $\tau_\ui^2 + \eta^2/4 > 1$, and $\tau_\up^2 +
\eta^2/4 > 1$. 
\subsubsection{Infinite time limit}
\label{sec:inftime}
If one further assumes that $\eta/\tau, \eta/\abs{\tau_\ui}, \eta/\tau_\up \ll 1$, then
$\cos[\theta(\tau)]$ and $\sin[\theta(\tau)]$ can be expanded in powers of $\eta/\tau$. We find
$\cos[\theta(\tau)] = 1 + \mathcal{O}(\eta^2/\tau^2)$ and $\sin[\theta(\tau)] =
\mathcal{O}(\eta/\tau)$. In this limit, which we refer to as the infinite-time limit, the return
probability becomes 
\begin{equation}
	P_{1\to 1}^\mathrm{inf} = 1 - 4 P_{\mm{LZ}} \left(1-P_{\mm{LZ}}\right) \cos^2 \left[\chi_{\mm{dp}} (\tau_\up) \right] 
	+ \mathcal{O}\left(\frac{\eta}{\tau}, \frac{\eta}{\tau_\ui}, \frac{\eta}{\tau_\up} \right),
	\label{eq:asymptrprob}
\end{equation}
where $P_{\mm{LZ}} = \lim_{t_\ui \to -\infty,\,t \to \infty} \abs{\varphi_{11} (t,t_\ui)}^2 =
\exp(-\pi \eta^2/2)$ is the Landau-Zener(-Stückelberg-Majorana) non-adiabatic transition
probability~\cite{1932_Landau,1932_Zener,1932_Stuckelberg,1932_Majorana} and we have defined the phase acquired during the double passage
\begin{equation}
	\chi_{\mm{dp}} (\tau) = -\frac{\eta^2}{4}+\frac{\eta^2}{2}
	\log\left[\frac{1}{2}\left(\tau + \sqrt{\tau^2 + \eta^2}\right)\right] 
	+ \frac{\tau}{2}\sqrt{\tau^2 + \eta^2} - \mm{arg}\left[\Gamma\left(1
	+i\frac{\eta^2}{4}\right)\right] - \frac{\pi}{4}.
	\label{eq:argcosP11}
\end{equation}

As we will show below, Eq.~\eqref{eq:asymptrprob} can also be found using the so-called
\textit{adiabatic impulse model}. While the latter model allows one to easily find an expression of
the return probability in the limit $\eta/\tau, \eta/\abs{\tau_\ui}, \eta/\tau_\up \ll 1$, it is
very hard to extend the adiabatic impulse model to other parameter regimes. Another drawback of the
adiabatic impulse model is that the leading order corrections to Eq.~\eqref{eq:asymptrprob} cannot
be found. In appendix~\ref{sec:returnPloc}, we give an expression for the leading order correction
to Eq.~\eqref{eq:asymptrprob}, which demonstrates that even in the infinite-time limit the return
probability depends explicitly on $\tau_\ui$ and $\tau$.

\subsection{Adiabatic impulse model}
\label{st}

In this section, we briefly recapitulate a previous theoretical approach to Stückelberg
interferometry known as the \textit{adiabatic impulse model}~\cite{2010_Shevchenko_Review}. The main
assumptions of the adiabatic impulse model are that all non-adiabatic transitions happen at $\tau=0$
and that the system follows perfect adiabatic evolution from $\tau_\ui \to 0_-$ and from $0_+ \to
\tau_\up$, where $\abs{\tau_\ui} \gg \eta$ with $\tau_\ui < 0$ and $\tau_\up  \gg \eta$.  Given the
assumptions of the model, it is convenient to work in the basis of instantaneous eigenstates of
Eq.~\eqref{eq:c1c2_matrix_lzsm}. 

The non-adiabatic part of the evolution is described with a scattering matrix that relates the
probability amplitudes right before the avoided crossing at $t=0_-$ and right after the avoided
crossing at $t=0_+$. The scattering matrix (in the basis of instantaneous eigenstates)
reads~\cite{2010_Shevchenko_Review}
\begin{equation}
	N = 
	\begin{pmatrix}
		\sqrt{1-P_{\mm{LZ}}} e^{-i (\chi_\us - \frac{\pi}{2})} && -\sqrt{P_{\mm{LZ}}} \\
		\sqrt{P_{\mm{LZ}}} && \sqrt{1-P_{\mm{LZ}}} e^{i (\chi_\us - \frac{\pi}{2})}
	\end{pmatrix}.
	\label{eq:Nmatrix}
\end{equation}
Here, $\chi_\us = \pi/4 + (\eta^2/4)[\log(\eta^2/4) -1] - \mm{arg}[\Gamma(1+i \eta^2/4)]$ is the
so-called Stokes phase~\cite{meyer1989}.

The adiabatic part of the evolution is described by the unitary evolution operator 
\begin{equation}
	U_{\mm{ad}} (\tau, \tau_\ui) = \exp\left[- i \chi_{\mm{dyn}} (\tau, \tau_\ui) \sigma_z\right],
	\label{eq:adiabaticevolution}
\end{equation}
with $\sigma_z$ the Pauli matrix in the z-direction and we have defined the dynamical phase 
\begin{equation}
	\begin{aligned}
	\chi_{\mm{dyn}} (\tau, \tau_\ui) 
	&= 
	\frac{1}{2} \int_{\tau_\ui}^\tau \ud \tau_1 \sqrt{\tau_1^2 + \eta^2} \\
	&= 
	\frac{1}{4}\left( \tau \sqrt{\tau^2 + \eta^2} + \eta^2
	\log\left[\tau + \sqrt{\tau^2 + \eta^2}\right] - \tau_\ui \sqrt{\tau_\ui^2 + \eta^2} - \eta^2
	\log\left[\tau_\ui + \sqrt{\tau_\ui^2 + \eta^2}\right] \right).
	\end{aligned}
	\label{eq:chidyn}
\end{equation}

Within this formalism, the state of the system after a double passage is given by 
\begin{equation}
	\mathbf{c}_{\mm{LZ}} (\tau) \simeq U_{\mm{ad}} (\tau, -0_-) N U_{\mm{ad}} (-0_+, -\tau_\up)
	U_{\mm{ad}} (\tau_\up, 0_+) N U_{\mm{ad}} (0_-, -\abs{\tau_\ui}) \mathbf{c}_{\mm{LZ}} (-\abs{\tau_\ui}).
	\label{eq:AIMstate}
\end{equation}
Here, we have chosen $0_-$ and $0_+$ to represent fixed times along the time axis. Note that we employ the scattering matrix $N$ instead of its Hermitian conjugate in the back sweep. Since the scattering matrix is expressed in the basis of instantaneous eigenstates, there is no difference in which direction the non-adiabatic transition is performed.

In general, all four adiabatic evolution operators in Eq.~\eqref{eq:AIMstate} contribute to the
acquired dynamical phase of the system. For the particular case of the presented experiment, we
initialize the system in an eigenstate of the coupled system, which is the out-of-plane mode. In
this scenario, the first and the last adiabatic evolution operators in Eq.~\eqref{eq:AIMstate} turn
into global phases, which do not contribute to the two-mode interference. Hence, the interference of
the two modes is solely governed by the phase evolution in between the two scattering matrices.
Finally, the adiabatic impulse model yields the return probability 
\begin{equation}
	P_{1\to 1}^\mathrm{aip} = 1-4 P_{\mm{LZ}} (1 - P_{\mm{LZ}}) \cos^2\left[\chi_{\mm{dp}}
	(\tau) \right],
	\label{eq:adimp}
\end{equation}
where we have used the fact that in the infinite-time limit ($\eta/\tau \ll 1$) the instantaneous
eigenstates and diabatic states of Eq.~\eqref{eq:c1c2_matrix_lzsm} coincide with each other.

As mentioned earlier, we have $P_{1\to 1}^\mathrm{aip} = P_{1\to 1}^\mathrm{inf}$. The main difficulty in using
the adiabatic impulse model to get the return probability in regimes other than $\eta/\tau \ll 1$ lies in
finding an appropriate scattering matrix $N$ that explicitly depends on time.

\section{Comparison of experiment and theories}
\label{exp}
\subsection{St\"uckelberg oscillations}
\label{sec:st}
\begin{figure}[!htb]
\includegraphics{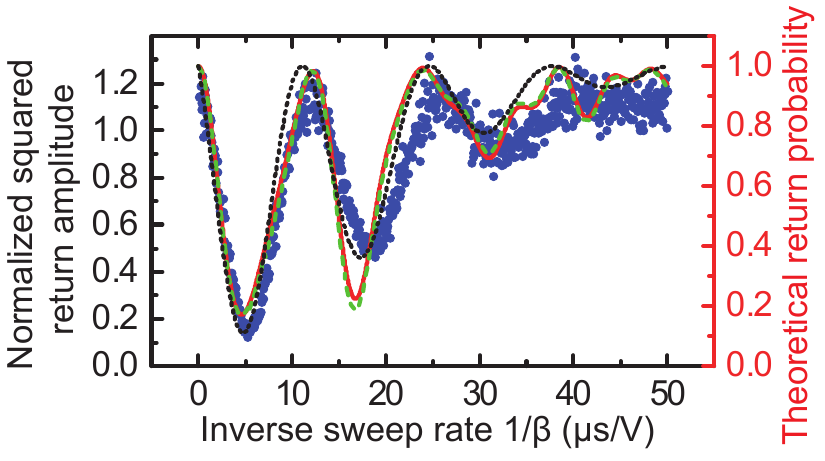}
\caption{\label{Fig_3} Classical St\"uckelberg interfermoetry measured on sample\,A. Normalized squared return amplitude versus inverse sweep rate $1/\beta$ for a peak voltage of $U_\mathrm{p}=2.5$\,V measured on sample\,A (left axis, blue dots). Right axis (red): Comparison of the different theoretical models calculated with a single set of parameters extracted from the experimental data; exact solution via finite times (red solid line), asymptotic theory (green dashed line) and adiabatic impulse model (black dotted line).}
\end{figure}

In this chapter, we compare the experimental results, taken with two different samples, with the
theoretical models from chapter\,\ref{theo}. The measurements on sample A are
performed at a temperature of 10\,K in a pulse tube cryostat. The 50\,\textmu m long nanomechanical
string resonator exhibits a linewidth of the mechanical resonance $\Gamma/2\pi\approx 40$\,Hz at the
out-of-plane resonance frequency $\omega_1(U_\mathrm{i})/2\pi = 7.560$\,MHz and hence a quality
factor $Q=\omega_1/\Gamma\approx 2\times 10^5$. Here, $U_\mathrm{i}=7.9$\,V is the initialization
voltage. The two strongly coupled mechanical modes exhibit a frequency splitting of $\Delta/2\pi =22.614$\,kHz at the avoided crossing voltage $U_\mathrm{a}= U_\mathrm{i}+1.471$\,V$=9.731$\,V. The
conversion factor from frequency sweep rate $\alpha$ to voltage sweep rate $\beta$ is determined
from the avoided crossing via a linear fit\cite{2016_Seitner_NatPhys} as $\zeta =55.042$\,kHz/V.
Figure\,\ref{Fig_3} depicts the measured normalized squared return amplitude versus inverse voltage
sweep rate $1/\beta$ for a fixed peak voltage $U_\mathrm{p}=2.5$\,V and read-out voltage
$U_\mathrm{f}=U_\mathrm{i}+0.2$\,V$=8.1$\,V. The experimentally taken data (blue points) is compared to the
different theoretical predictions calculated from the above parameters. The red solid line
represents the exact solution following Eq.~\eqref{eq:returnprob}, whereas the green dashed line depicts
the asymptotic theory (see appendix\,\ref{app:asymptotic}) and the black dotted line shows the results from the
adiabatic impulse model [Eq.~\eqref{eq:adimp}].
We find the experimental data in good agreement with the exact solution via finite times and the asymptotic theory, where the latter nearly coincides with the exact finite time solution. Explicitly, the accumulation of phase and hence the state interference is determined by the exact evolution of the two frequency branches in time in combination with the coupling of the modes. This dependence can be absorbed in characteristic times and an effective coupling $\eta$, determined by the sweep rate and the voltages of initialization, avoided crossing, turning-point and read-out:
\begin{eqnarray}
\label{eq:conv_time}
\begin{aligned}
&t_\mathrm{i}=-\frac{1}{\beta}(U_\mathrm{a}- U_\mathrm{i})=\frac{\tau_\mathrm{i}}{\sqrt{\alpha}}\\
&t_\mathrm{p}=\frac{1}{\beta}(\widetilde{U}_\mathrm{p}- U_\mathrm{a})=\frac{\tau_\mathrm{p}}{\sqrt{\alpha}}\\
&t_\mathrm{f}=\frac{1}{\beta}(U_\mathrm{a}- U_\mathrm{f})=\frac{\tau_\mathrm{f}}{\sqrt{\alpha}}\\
&\eta=\frac{\Delta}{\sqrt{2\pi\zeta\beta}}=\frac{\Delta}{\sqrt{\alpha}}.
\end{aligned}
\end{eqnarray}
Note that in the definition of $t_\mathrm{p}$, the absolute peak voltage $\widetilde{U}_\mathrm{p}$ appears instead of the peak voltage of the applied voltage ramp $U_\mathrm{p}$ (see Fig.\,\ref{Fig_2}\,b).\\
The St\"uckelberg return probability predicted by the adiabatic impulse model exhibits good agreement with the experimental data as well. Nevertheless, distinct features appearing in the exact finite time solution and the asymptotic theory are not reproduced by the adiabatic impulse model. This can be understood as follows: From the experimental parameters in Fig.\,\ref{Fig_3} we find the system to operate in the limit of \textit{infinite times} (cf. chapter\,\ref{sec:inftime}) since $\eta/\tau_\mathrm{f}\approx 0.05, \eta/\abs{\tau_\mathrm{i}}\approx 0.04, \eta/\tau_\mathrm{p}\approx 0.06 \ll 1$. As detailed in appendix\,\ref{sec:returnPloc}, the adiabatic impulse model corresponds to the zeroth order series expansion of $\cos[\theta(\tau)]$ and $\sin[\theta(\tau)]$ in the long time limit of the asymptotic solution. Taking into account higher order corrections to the return probability would result in the appearance of the distinct features of the exact solution in the extended adiabatic impulse model. \\
\begin{figure*}[!htb]
\includegraphics{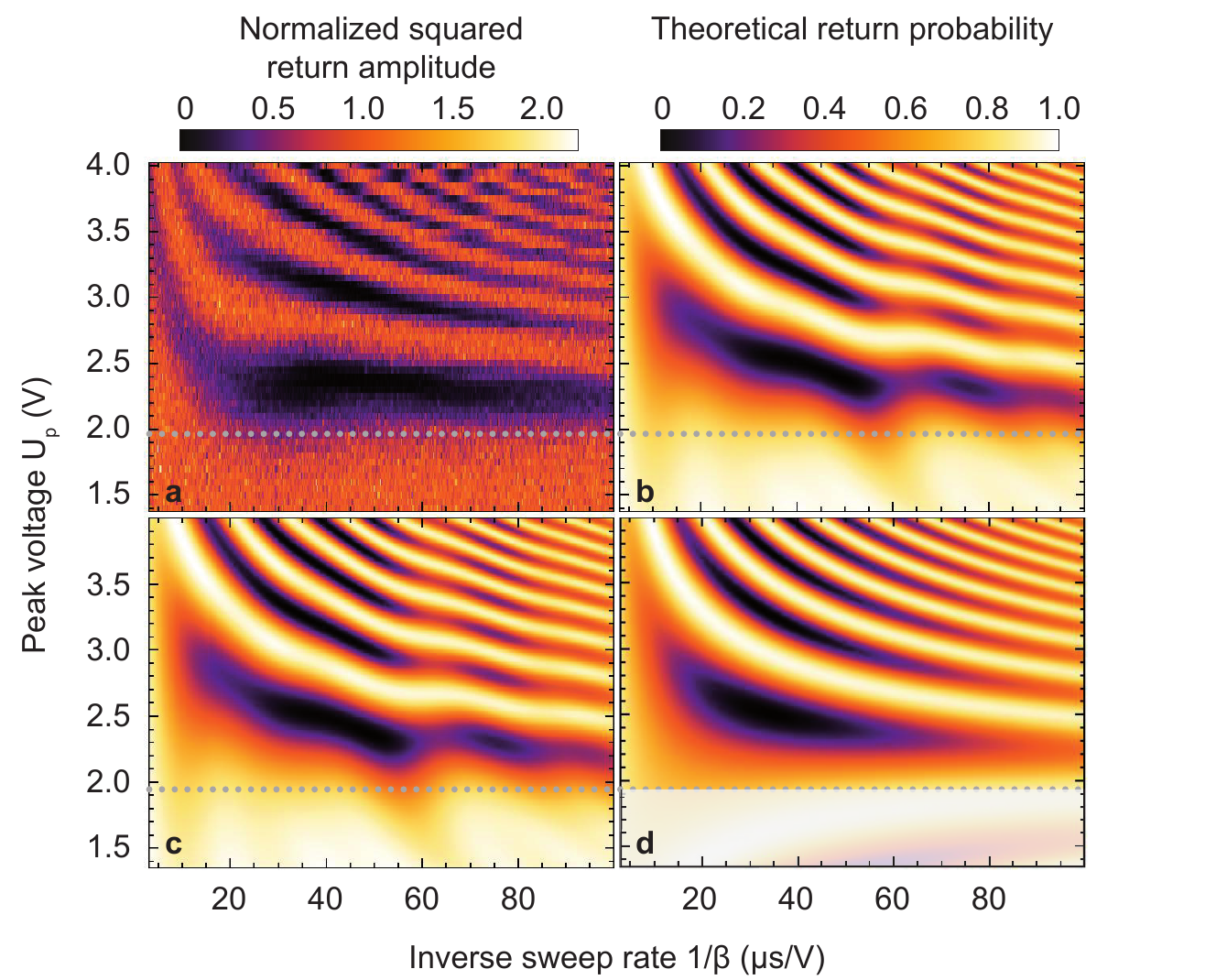}
\caption{\label{Fig_4} Comparison of experiment and different theoretical models on St\"uckelberg interferometry measured on sample\,B. \textbf{a},\,Color-coded (left colorbar) normalized squared return amplitude versus peak voltage $U_\mathrm{p}$ and inverse sweep rate $1/\beta$. Gray dotted line indicates the position of the avoided crossing at voltage $U_\mathrm{p}=1.958$\,V in all sub-panels. \textbf{b}-\textbf{d},\,Color-coded theoretical return probability (right colorbar) calculated from a single set of parameters, extracted from the experimental data, for the exact finite time solution\,(\textbf{b}), the asymptotic theory\,(\textbf{c}) and the adiabatic impulse model\,(\textbf{d}). The mechanical damping is modelled according to Eq.~\eqref{eq:c_damp} in all theory plots (\textbf{b}-\textbf{d}). Note that the adiabatic impulse model (\textbf{d}) is plotted for the same parameter range without respect to the physical validity in certain ranges. The region for peak voltages $\widetilde{U}_\mathrm{p} < U_\mathrm{B,a}$ is manually grayed out as explained in the text.}
\end{figure*}
In order to investigate the validity of the different theoretical approaches, we perform St\"uckelberg interferometry experiments for a large set of peak voltages $U_\mathrm{p}$ at room temperature using a second sample. In particular, we study the finite time dynamics of the system for absolute peak voltages, i.e. turning points, close to the avoided crossing and even observe interference without traversing the latter. The sample measured at room temperature (sample B, denoted "B") incorporates the same basic design as sample\,A except the length of the nanomechanical string resonator of 55\,\textmu m. However, sample\,B exhibits a mechanical resonance linewidth of $\Gamma_\mathrm{B}/2\pi\approx 25$\,Hz at frequency $\omega_\mathrm{B,1}(U_\mathrm{B,i})/2\pi=6.561$\,MHz and consequently provides an enhanced mechanical lifetime of 6.21\,ms. The system is initialized at $U_\mathrm{B,i}=10.4$\,V, again in the lower frequency branch, and read-out at $U_\mathrm{B,f}=U_\mathrm{B,i}+0.5$\,V$=10.9$\,V. The two modes exhibit a frequency splitting of $\Delta_\mathrm{B}/2\pi=6.322$\,kHz at the avoided crossing voltage $U_\mathrm{B,a}=U_\mathrm{B,i}+1.958$\,V$=12.358$\,V and the conversion factor is determined as $\zeta_\mathrm{B}=19.224$\,kHz/V. Figure\,\ref{Fig_4}\,a depicts a color-coded two-dimensional map of the normalized squared return amplitude in dependence of the inverse voltage sweep rate $1/\beta$ and the peak voltage $U_\mathrm{p}$. The gray dotted line indicates the avoided crossing voltage $U_\mathrm{B,a}$. We observe clear interference fringes in the normalized squared return amplitude even for peak voltages near or smaller than the avoided crossing voltage ($\widetilde{U}_\mathrm{p}\lesssim U_\mathrm{B,a}$). This fact proves that the two modes interfere without the explicit need of traversing the avoided crossing and clearly indicates the importance of finite time dynamics in St\"uckelberg interferometry.\\
Calculating the theoretical return probability using the exact solution via finite time dynamics from Eq.~\eqref{eq:returnprob} with no free parameters yields good agreement between experiment and theory as displayed in Fig.\,\ref{Fig_4}\,b. Because of the longer duration ramps applied to sample\,B (up to $\vartheta=1.0$\,ms), the mechanical damping needs to be taken into account. Modelling the mechanical damping by an exponential decay with average energy decay time\cite{2016_Seitner_NatPhys} $t_0=5.7$\,ms, the evolution of the modes after the double sweep through the avoided crossing is given by
\begin{equation}
\label{eq:c_damp}
|c_{j}(t)|^2=\exp[-t/t_0]P_{1\to j},
\end{equation}
with $P_{1\to j}$ the return probability to mode $j=1$ or $j=2$. Note that Eq.~\eqref{eq:c_damp} is applied to all three different theoretical approaches in the following.\\
In Fig.\,\ref{Fig_4}\,b, the self-interference of the two-mode system clearly extends beyond the dotted line which represents the position of the avoided crossing at voltage $U_\mathrm{B,a}$. Additionally, the theory exhibits distinct features in the interference pattern which are not reproduced by the experiment. In order to experimentally resolve these features, the system needs to interfere precisely with the same constant parameters in every particular measurement pixel from Fig.\,\ref{Fig_4}\,a. Since the experiments are performed at room temperature, the system parameters vary strongly from measurement to measurement due to temperature fluctuations. Experimentally, we partially account for this effect by the implementation of an initialization voltage feedback loop\cite{2016_Seitner_NatPhys}, which ensures the initialization of the system at the same resonance frequency, at least within one horizontal line from Fig.\,\ref{Fig_4}\,a. Nevertheless, the fluctuations and uncertainties prevent the system from interfering with the precise same parameters in every particular measurement. Further information on the experimental uncertainties is provided elsewhere\cite{2016_Seitner_NatPhys}. Note that the experimental data in Fig.\,\ref{Fig_4}\,a is taken in a non-consecutive way over a timespan of approximately 6 months which clearly demonstrates the validity of the data.\\
The results for the asymptotic theory (appendix\,\ref{app:asymptotic}) are depicted in Fig.\,\ref{Fig_4}\,c. The asymptotic theory reproduces the exact finite time solution with excellent agreement over the full displayed parameter range. Consequently, one can exploit the piecewise definition of the asymptotic theory in appendix\,\ref{app:asymptotic} to deduce the characteristic dynamics of the system in each particular parameter regime. A detailed discussion of the different regimes will be given in chapter\,\ref{sec:regimes}.\\
Figure\,\ref{Fig_4}\,d depicts the return probability calculated from the adiabatic impulse model [Eq.~\eqref{eq:adimp}] using the same parameters as for the exact solution via finite times. The model is depicted for the complete experimentally investigated parameter regime. However, certain displayed parameter ranges flaw the basic assumptions of the adiabatic impulse model that the dynamics of the system is fully adiabatic and governed by ``infinite times" as described in chapter\,\ref{st}. Since there is no sharp transition for the validity of the model, the full parameter range is displayed. Additionally, the region below the avoided crossing voltage (gray dotted line) in Fig.\,\ref{Fig_4}\,d is grayed out manually. The reason is that Eq.~\eqref{eq:adimp} is unphysical for the region where $\widetilde{U}_\mathrm{p} < U_\mathrm{B,a}$ since the definition of the adiabatic impulse model requires traversing the avoided crossing.\\
In general, we observe a clear deviation between the adiabatic impulse model and both, the experimental data and the exact solution. In particular, the interference fringes of Fig.\,\ref{Fig_4}\,d vanish for peak voltages in the region of the avoided crossing, when the dimensionless time of the phase evolution $\tau_\mathrm{p}$ becomes comparable to the dimensionless level splitting $\eta$. This discrepancy clearly demonstrates that the dynamics of the system cannot be generally described by an infinite time approach where the two-mode interference is solely governed by the coupling of the system. However, for peak voltages, i.e., turning points far away from the avoided crossing, the result of the adiabatic impulse model qualitatively resembles the result obtained by the exact solution.
\subsection{Interference visibility}
\label{sec:vis}
\begin{figure*}[!htb]
\includegraphics{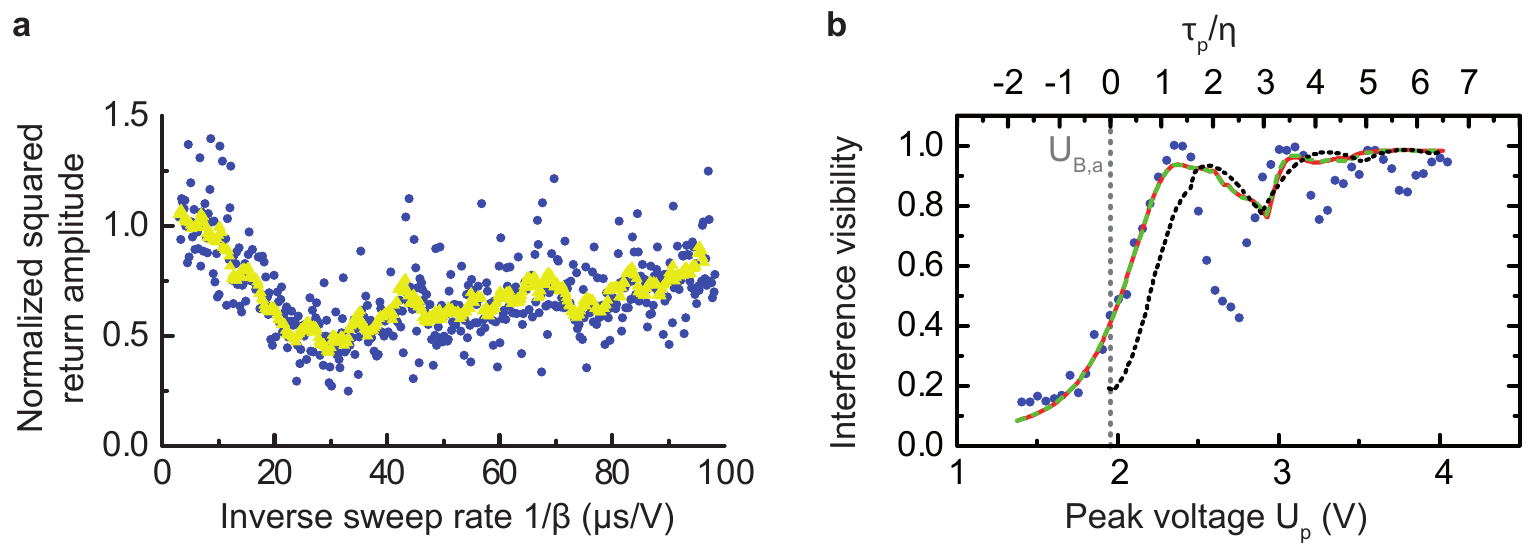}
\caption{\label{Fig_5} Interference visibility. \textbf{a},\,Exemplary line-cut taken along the gray dotted line in Fig.\,\ref{Fig_4}\,a for illustration of the averaging of the experimental data. Blue dots represent the experimentally determined normalized squared amplitude versus inverse sweep rate $1/\beta$ for peak voltage $\widetilde{U}_\mathrm{p} \approx U_\mathrm{B,a}$. Yellow triangles correspond to the averaged data using a moving average of 10 points. \textbf{b},\,Interference visibility as a function of peak voltage $U_\mathrm{p}$ (bottom axis) and characteristic dimensionless ratio $\tau_\mathrm{p}/\eta$ (top axis) extracted from horizontal line-cuts of Fig.\,\ref{Fig_4}\,b-d for the exact theoretical solution (red solid line), asymptotic theory (green dashed line) and adiabatic impulse model (black dotted line). Blue dots depict the interference visibility calculated from the \textit{averaged} experimental data (yellow triangles in panel\,\textbf{a}) as described in the text.}
\end{figure*}
In order to study the crossover from the infinite time limit to the finite time domain in more
detail, we extract the interference visibility in dependence of the peak voltage from the
experimental data and the different theoretical approaches. Note that by interference visibility we
refer to the original definition of interference contrast\cite{Saleh_Teich} and not to the
single-shot read-out-visibility as frequently referred to in e.g. spin
systems\cite{2010_Morello_Nature}. The interference visibility from the experimental data for a
given peak voltage is calculated from the corresponding horizontal line-cut in Fig.\,\ref{Fig_4}\,a
by the difference of the maximum and minimum normalized squared return amplitude divided by their
sum\cite{Saleh_Teich}. As exemplarily illustrated in Fig.\,\ref{Fig_5}\,a, we used a moving average
of 10 points (see appendix\,\ref{ma}) in each respective horizontal line to account for experimental
scatter of the data in the calculation of the interference visibility. The dataset corresponds to
the line-cut along the gray dotted line in Fig.\,\ref{Fig_4}\,a where $\widetilde{U}_\mathrm{p}
\approx U_\mathrm{B,a}$. Figure\,\ref{Fig_5}\,b depicts the interference visibility as a function of
peak voltage for the experimental data (blue dots) and the different theoretical models. Since the
theoretical models represent real probabilities, we associate the interference visibility with the
interference contrast, which is the difference of the maximum and the minimum return probability
without a normalization to their sum. The exact solution via finite times [red solid line,
Eq.~\eqref{eq:returnprob}] clearly exhibits a non-zero interference visibility for a set of peak
voltages smaller than the avoided crossing voltage $U_\mathrm{B,a}$ (gray dotted vertical line) in
good agreement with the experimental data. The interference visibility obtained from the return
probability in the asymptotic theory (green dashed line) nearly coincides with the result of the
exact finite time solution. The underlying agreement clearly demonstrates that the exact solution of
the double passage St\"uckelberg problem via finite times can be well-approximated by taking the
asymptotic limit of the parabolic cylinder functions in the appropriate parameter regime.
Furthermore, the analysis confirms the appearance of interference for peak voltages before the
avoided crossing. In contrast, the adiabatic impulse model [black dotted line, Eq.~\eqref{eq:adimp}]
interference visibility drops down close to zero for peak voltages smaller than $U_\mathrm{p}\approx
2.0$\,V, such that $\widetilde{U}_\mathrm{p} \approx U_\mathrm{B,a}$.

To obtain a more intuitive understanding of the interference visibility, we replace the parameter $U_\mathrm{p}$ by introducing
the ratio between two characteristic dimensionless scales, which are the dimensionless time
$\tau_\mathrm{p}$ and the dimensionless coupling $\eta$ (see top x-axis in Fig.\,\ref{Fig_5}\,b). As explained above
[Eqs.~\eqref{eq:conv_time}], $\tau_\mathrm{p}$ corresponds to the distance from the avoided
crossing to the turning point of the double sweep and hence becomes negative for sweeps where the
avoided crossing is not passed (cf. top axis in Fig.\,\ref{Fig_5}\,b). The dimensionless coupling
$\eta$ represents the effective level splitting between the two modes and is itself independent of
the turning point in the double sweep [cf. Eqs.~\eqref{eq:conv_time}]. Consequently, a
characteristic dimensionless ratio of $\tau_\mathrm{p}/\eta=1$ would correspond to a population
transfer from the lower to the upper mode with fidelity of 100\,\% using the generic picture of the
Bloch sphere\cite{Nielsen2000, Dragoman2004} in the classical two-mode
system\cite{2013_Faust_NatPhys}. Losely speaking, the system has enough "time" to perform a complete
population transfer to the upper mechanical mode when initialized in the lower mode. In principle,
this characteristic behaviour can be extracted from the interference visibility depicted in
Fig.\,\ref{Fig_5}\,b. For $\tau_\mathrm{p}/\eta=1$ the interference visibility of the experimental
data (blue dots) and the exact theoretical finite time solution (red solid line) reaches a maximum
which is close to unity. At this point, we recover the full interference contrast since the two
modes have the ability to interfere fully destructive due to the possibility of a complete
population transfer. However, the interference visibility extracted from the theory saturates to a
value of approximately 94\,\% whereas the experimental data converges to 100\,\% visibility. The
origin of this discrepancy in the theory is attributed to the fact that the interference pattern in
Fig.\,\ref{Fig_4}\,b is calculated for the parameter range in which the experiment is conducted. The
fastest voltage sweeps are performed at an inverse sweep rate of $1/\beta=3$\,\textmu s/V. Whereas
experimental scatter of the data allows for a visibility of 100\,\%, the theory does not incorporate
sufficiently fast sweeps to return to the same mode with a probability of unity, i.e., the sweeps
are not non-adiabatic enough. In the limit $1/\beta\to 0$, the theory would simultaneously exhibit a
100\,\% visibility at $\tau_\mathrm{p}/\eta=1$. For $\tau_\mathrm{p}/\eta\gg 1$ the theory curve
also converges to 100\,\% visibility since the horizontal line-cuts from Fig.\,\ref{Fig_4} intersect
at least one constructive and one destructive interference fringe. Furthermore, we observe a reduced
interference visibility in certain regions, where the return probability does not completely drop
down to zero. We attribute this to the hyperbolic shape of the observed interference fringes in the
displayed parameter space representation of the return probability as a function of inverse sweep
rate and peak voltage. One could easily find horizontal line-cuts in Fig.\,\ref{Fig_4} where the
return probability does not completely drop down to zero, which translates into a reduced
interference visibility. In this work, it is not investigated further if there is a distinct
physical origin of this reduction in interference visibility.

In contrast to the above findings, the interference visibility extracted from the adiabatic impulse
model (black dotted line) peaks for a larger ratio of $\tau_\mathrm{p}/\eta\approx 1.5$,
which clearly demonstrates that this model is only valid if the turning point is far away from the
avoided crossing. However, we recover qualitatively similar dips in the interference visibility as in the
exact finite time solution and the experimental data.\\
For larger ratios of $\tau_\mathrm{p}/\eta$, i.e. $\eta/\tau_\mathrm{p} \ll 1$, the interference visibility extracted from the adiabatic impulse model coincides with the exact finite time solution. This result is in excellent agreement with the definition of the adiabatic impulse model as the infinite time limit of the finite time St\"uckelberg theory [cf. section\,\ref{sec:inftime}].

\subsection{Parameter regimes}
\label{sec:regimes}
In this section, we exploit the piecewise definition of the asymptotic limit of the exact theoretical finite time solution given in appendices\,\ref{app:asymptotic}\,\&\,\ref{sec:returnPloc} to quantify specific parameter regimes of St\"uckelberg interferometry. Depending on the specific regime, the characteristic times [Eqs.~\eqref{eq:conv_time}] are limited to certain boundaries. These boundaries, in turn, allow for the quantification of the underlying physics governing the coupled system dynamics.

In order to asymptotically expand the parabolic cylinder functions, we define a critical
dimensionless time $\tau_\mathrm{crit}$. This critical dimensionless time serves as a measure for
the employed dimensionless times $\tau_\mathrm{i}$, $\tau_\mathrm{p}$ and $\tau_\mathrm{f}$ in the
theory. Those parameters can either be bound by $\tau_\mathrm{crit}$ ($-\tau_\mathrm{crit}\leq
\tau\leq \tau_\mathrm{crit}$) or unbound ($\abs{\tau} > \tau_\mathrm{crit}$). Here, one should keep in mind that $\tau_\mathrm{i}$ is defined as smaller than zero. As further discussed
in appendix\,\ref{app:asymptotic}, the parabolic cylinder functions can mathematically be
approximated by a power series. Hereby, the magnitude of $\tau_\mathrm{crit}$ specifies up to which
order the power series is expanded. For the following calculations we defined
$\tau_\mathrm{crit}=2$.

\begin{figure}[!htb]
\includegraphics{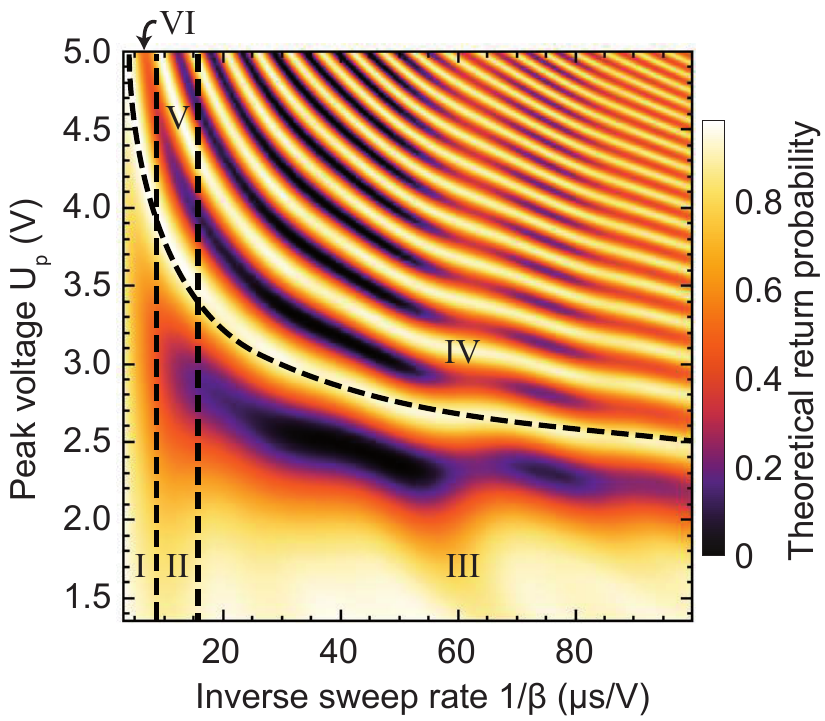}
\caption{\label{Fig_6} Phase space of the parameter regimes in the asymptotic theory. Color-coded theoretical return probability as calculated from the asymptotic theory (cf. Fig.\,\ref{Fig_4}\,c) for an extended peak voltage range. Black dashed lines indicate the border-lines of the different parameter regimes in the asymptotic theory. The different parameter regimes are labeled by roman numerals which are elucidated in Table\,\ref{tab:regimes}.}
\end{figure}

Figure\,\ref{Fig_6} depicts the theoretical return probability calculated from the asymptotic theory as in Fig.\,\ref{Fig_4}\,c for an extended peak voltage range. The layover in Fig.\,\ref{Fig_6} displays the boundary lines of the different parameter regimes from which the asymptotic solution is calculated. We recover six different parameter regimes, labeled by roman numerals from I to VI, which are specified in Table\,\ref{tab:regimes}.
\begin{table}
\begin{tabular}{l*{3}{c}}
Regime              & $\tau_\mathrm{i}$ & $\tau_\mathrm{p}$ & $\tau_\mathrm{f}$ \\
\hline
\hline
I      & $-\tau_\mathrm{crit}\leq \tau_\mathrm{i}\leq \tau_\mathrm{crit}$ & $-\tau_\mathrm{crit}\leq \tau_\mathrm{p}\leq \tau_\mathrm{crit}$ & $-\tau_\mathrm{crit}\leq \tau_\mathrm{f}\leq \tau_\mathrm{crit}$ \\
II     & $\tau_\mathrm{i}< -\tau_\mathrm{crit}$ &  $-\tau_\mathrm{crit}\leq \tau_\mathrm{p}\leq \tau_\mathrm{crit}$ & $-\tau_\mathrm{crit}\leq \tau_\mathrm{f}\leq \tau_\mathrm{crit}$ \\
III    & $\tau_\mathrm{i} < -\tau_\mathrm{crit}$ & $-\tau_\mathrm{crit}\leq \tau_\mathrm{p}\leq \tau_\mathrm{crit}$ & $\tau_\mathrm{f} > \tau_\mathrm{crit}$ \\
IV     & $\tau_\mathrm{i} < -\tau_\mathrm{crit}$ & $\tau_\mathrm{p} > \tau_\mathrm{crit}$ & $\tau_\mathrm{f} > \tau_\mathrm{crit}$ \\
V      & $\tau_\mathrm{i} < -\tau_\mathrm{crit}$ & $\tau_\mathrm{p} > \tau_\mathrm{crit}$ & $-\tau_\mathrm{crit}\leq \tau_\mathrm{f}\leq \tau_\mathrm{crit}$ \\
VI     &  $-\tau_\mathrm{crit}\leq \tau_\mathrm{i}\leq \tau_\mathrm{crit}$ & $\tau_\mathrm{p} > \tau_\mathrm{crit}$ & $\tau_\mathrm{f} > \tau_\mathrm{crit}$ \\
\hline
\hline
\end{tabular}
\caption{\label{tab:regimes}Summary of the different asymptotic regimes.}
\end{table}
As can be seen immediately from Table\,\ref{tab:regimes}, there is only one regime where all three characteristic times are above threshold, which is regime IV. Since the characteristic times are not bound in this regime it is considered as the long time limit, which includes the infinite time limit, e.g. the adiabatic impulse model, where $\eta/\tau_\mathrm{f}, \eta/\abs{\tau_\mathrm{i}}, \eta/\tau_\mathrm{p} \ll 1$. However, there is no sharp border between the long time limit and the infinite time limit. Whereas the former requires the dimensionless times to be much larger than one, the latter exhibits the additional constraint that the dimensionless times are large compared to the dimensionless coupling. In this infinite time limit, the dynamics of the strongly coupled two-mode system is governed by the coupling strength of the two modes since the exact evolution in terms of dimensionless time plays a minor role. In fact, this is the only regime which, to the best of our knowledge, has been considered in the past in the framework of Landau-Zener type physics\cite{1932_Landau,1932_Zener,2010_Shevchenko_Review} and St\"uckelberg interferometry\cite{2010_Shevchenko_Review}, except for the work of \textit{Vitanov et al.}\cite{1996_Vitanov_PRA} and Refs.\,\onlinecite{2013_Ribeiro_PRL,2013_Ribeiro_PRB}. Nevertheless and as one can easily deduce from Fig.\,\ref{Fig_6}, a complete solution of the double passage St\"uckelberg problem is in crucial need of additional parameter regimes, where the finite durations of the sweeps play a major role.\\
The transition from the long time limit to the finite time domain is represented by the hyperbolic black dashed line in Fig.\,\ref{Fig_6}. Associating a threshold peak voltage $U_\mathrm{p,crit}$ with this transition, one can easily calculate the border-line as a function of inverse sweep rate via the definitions of $\tau_\mathrm{p}$ [cf. Eqs.~\eqref{eq:conv_time}] and $\tau_\mathrm{crit}=2$
\begin{eqnarray}
\label{eq:ucrit}
\begin{aligned}
U_\mathrm{p,crit}&= \sqrt{\frac{2}{\pi\zeta}}\frac{1}{\sqrt{1/\beta}}+(U_\mathrm{B,a}-U_\mathrm{B,i})\\
                 &\approx \sqrt{\frac{2}{\pi\zeta}}\frac{1}{\sqrt{1/\beta}}+1.958\,\mathrm{V}.
\end{aligned}
\end{eqnarray}
Accordingly, the two vertical border-lines in Fig.\,\ref{Fig_6}, which are independent of the peak voltage ($[1/\beta]_\mathrm{I}=8.64$\,\textmu s/V, $[1/\beta]_\mathrm{II}=15.58$\,\textmu s/V), can be calculated straight forward from Eqs.~\eqref{eq:conv_time}. Note that for $\tau_\mathrm{i}=-\tau_\mathrm{f}$, i.e., if the system could be read-out at the initialization point after a symmetric voltage ramp, the right vertical border-line in Fig.\,\ref{Fig_6} would vanish. Even though regime II and regime V are exclusively observed in our particular measurement scheme, the importance of finite time effects in St\"uckelberg interferometry is definitely pointed out by the presence of regimes I, III and VI. Especially regime I and III are of great interest since they reveal the dynamics of St\"uckelberg interferometry between two strongly coupled modes without the explicit need of traversing the avoided energy level crossing.\\
Since the dynamics of the strongly coupled classical two-mode system can be mapped onto the dynamics of a quantum mechanical two-level system in St\"uckelberg interferometry\cite{2016_Seitner_NatPhys}, the same regimes are existent in every quantum mechanical two-level system such as e.g. superconducting qubits\cite{2005_Oliver_Science,2006_Silanpaa_PRL,2009_LaHaye_Nature,2012_Shevchenko_PRB} or spin-1/2 systems\cite{2010_Petta_Science,Gaudreau2012,2013_Ribeiro_PRL}. To the best of our knowledge, such regimes have so far not been investigated in the framework of St\"uckelberg interferometry and might be a prominent candidate for future investigations of quantum two-level systems.
\section{Conclusion}
\label{sec:conc}
In conclusion, we have demonstrated the importance of finite time effects in St\"uckelberg interferometry. Providing a complete and exact theoretical solution to the double passage St\"uckelberg problem, we have shown that the commonly employed adiabatic impulse model\cite{2010_Shevchenko_Review} does not address the full complexity of the problem\cite{1932_Stuckelberg}. In particular, the adiabatic impulse model solely describes one single parameter regime, where the dynamics of the system is completely governed by the coupling of the two modes corresponding to an infinite time limit. We have been able to asymptotically expand the provided exact finite time theoretical model and have hereby classified previously undiscovered parameter regimes in St\"uckelberg interferometry. The theoretical findings have been confirmed in remarkably good agreement by a detailed experimental study of the dynamics of a classical two mode system\cite{2013_Okamoto_NatPhys,2013_Faust_NatPhys} realized by two strongly coupled high quality factor nanomechanical string resonator modes. All theoretically predicted parameter regimes have been demonstrated experimentally by a thorough investigation of classical St\"uckelberg interferometry\cite{2016_Seitner_NatPhys}. We observed clear oscillations in the experimentally accessible normalized squared return amplitude, even without traversing the avoided crossing in excellent agreement with the provided exact theory. These findings have been supported by a detailed study of the interference visibility over a huge parameter range. Interestingly, the dynamics of the investigated classical two-mode system can be mapped to the dynamics of quantum mechanical two-level systems, as has recently been demonstrated by the authors\cite{2016_Seitner_NatPhys}. As a consequence, the above theoretical findings can be applied one-to-one to quantum mechanical two-level systems.

\begin{appendix}
\renewcommand\thefigure{\thesection.\arabic{figure}}
\section{Asymptotic expansion of the parabolic cylinder function}
\label{app:asymptotic}

In this section we list the asymptotic expansions used to produce Fig.\,\ref{Fig_4}\,c and Fig.\,\ref{Fig_6}. 

\subsection{Short-time expansion}

When $-\tau_{\mm{crit}} < \tau < \tau_{\mm{crit}}$, one can approximate parabolic cylinder functions by a
power series. In this work we used one of the power series derived in Ref.~\onlinecite{abadir1993},
\begin{equation}
	D_\nu (\tau) = \sqrt{\pi} 2^{\frac{\nu}{2}}
	\exp\left(\frac{\tau^2}{4}\right)\sum_{n=0}^\infty \frac{(-\sqrt{2} \tau)^n}{n!
		\Gamma\left[\frac{1}{2}(1-\nu-n)\right]}.
	\label{eq:seriesDnz}
\end{equation}
This expansion is particularly useful when $\abs{\tau_{\mm{crit}}} \ll 1$ since the series can be
truncated after a few terms. 

Here, we are going to choose $\abs{\tau_{\mm{crit}}} =2$. While we will not be able to truncate the
series to only one or two terms, we will be able to approximate the parabolic cylinder functions
with only two different functions. The special values $\abs{\tau_{\mm{crit}}} =2$ correspond then to
the point where the functions are matched.

\subsection{Long-time expansion}

When $\tau \gg 1$, one can use the results of Ref.~\onlinecite{olver1959} to find
asymptotic expansions for the relevant parabolic cylinder functions involved in Eqs.~\eqref{eq:phi11}
and \eqref{eq:phi12}. The asymptotic expansions are
\begin{equation}
	\begin{aligned}
		D_{-i \frac{\eta^2}{4} -1} \left(e^{i \frac{\pi}{4}} \tau\right)
		&\simeq 
		\frac{2}{\eta}\sin[\theta( \tau)] \exp\left[\frac{\pi \eta^2}{16} - i \left(\xi(\tau) +
		\frac{\pi}{4}\right)\right], \\
		D_{-i \frac{\eta^2}{4} -1} \left(e^{-i \frac{3\pi}{4}} \tau\right)
		&\simeq 
		\frac{2}{\eta}\sin[\theta( \tau)] \exp\left[-3\frac{\pi \eta^2}{16} - i
		\left(\xi(\tau) - \frac{3\pi}{4}\right)\right] \\
		&\phantom{={}} 
		+ \frac{\sqrt{2 \pi}}{\Gamma\left(1 + i \frac{\eta^2}{4}\right)}
		\cos[\theta( \tau)] \exp\left[-\frac{\pi \eta^2}{16} + i \xi(\tau)\right], \\
		D_{-i \frac{\eta^2}{4}} \left(e^{-i \frac{3\pi}{4}} \tau\right)
		&\simeq 
		\cos[\theta( \tau)] \exp\left[-3\frac{\pi \eta^2}{16} - i \xi(\tau)\right] \\
		&\phantom{={}}
		+\frac{\eta \sqrt{\pi}}{\sqrt{2} \Gamma\left(1 + i \frac{\eta^2}{4}\right)} 
		\sin[\theta( \tau)] %\times \\
		%&\phantom{={}}
		\exp\left[\frac{-\pi \eta^2}{16} + i \left(\xi(\tau) +
		\frac{\pi}{4}\right)\right], \\
		D_{-i \frac{\eta^2}{4}} \left(e^{i \frac{\pi}{4}} \tau\right)
		&\simeq
		\cos[\theta( \tau)] \exp\left[\frac{\pi \eta^2}{16} - i \xi(\tau)\right],
	\end{aligned}
	\label{eq:asymptoticsDnz}
\end{equation}
where we have defined 
\begin{equation}
	\begin{aligned}
		\sin[\theta( \tau)] &= \sqrt{\frac{1}{2} \left(1 - \frac{\tau}{\sqrt{\tau^2 +
		\eta^2}}\right)},\\
		\cos[\theta( \tau)] &= \sqrt{\frac{1}{2} \left(1 + \frac{\tau}{\sqrt{\tau^2 +
		\eta^2}}\right)},\\
		\xi (\tau) &= -\frac{\eta^2}{8} + \frac{\eta^2}{4}
		\log\left[\frac{1}{2}\left(\tau + \sqrt{\tau^2 + \eta^2}\right)\right] +\frac{\tau}{4}\sqrt{\tau^2 + \eta^2}.
	\end{aligned}
	\label{eq:defasymptoticsDnz}
\end{equation}

Note that this expansion is employed for $\tau \geq \tau_\mathrm{crit}$.\\
Finally, we would like to draw attention to the fact that Eq.~\eqref{eq:asymptoticsDnz} is also valid for
the weaker condition $\tau^2 + \eta^2/4 \gg 1$.

\subsubsection{``Negative'' long-time expansion}

To obtain the asymptotic expansions for negative arguments, $\tau < 0$ and $\abs{\tau} \gg 1$, one
substitutes $\tau \to e^{\pm i\pi} \abs{\tau}$ in the argument of the functions to be expanded. With
this substitution, the problem is reduced to the cases presented in Eqs.~\eqref{eq:asymptoticsDnz}. 

\subsubsection{Expansion of $[D_\nu(z)]^\ast$}

Since the parabolic functions are analytic, we have $[D_\nu (z)]^\ast = D_{\nu^\ast} (z^\ast)$. As a
consequence, the asymptotic expansion of $D_{\nu^\ast} (z^\ast)$ is the complex conjugate of the
asymptotic expansion of $D_\nu (z)$. 

\section{Leading order correction to the return probability in the infinite-time limit}
\label{sec:returnPloc}

As explained in the main text, we have defined the infinite-time limit as  $\eta/\tau,
\eta/\abs{\tau_\ui}, \eta/\tau_\up \ll 1$. To obtain the first-order correction to
Eq.~\eqref{eq:asymptrprob}, we use Eqs.~\eqref{eq:asymptphi11} and \eqref{eq:asymptphi12} and
expand $\cos[\theta(\tau)]$ and $\sin[\theta(\tau)]$ in powers of $\eta/\tau$. In contrast to what
has been presented in the main text, we keep the lowest contribution in $\eta/\tau$. We find
\begin{equation}
	\cos[\theta(\tau)] = 1 - \frac{1}{8}\frac{\eta^2}{\tau^2} +
	\mathcal{O}\left(\frac{\eta^3}{\tau^3}\right),
	\label{eq:coseta2}
\end{equation}
and
\begin{equation}
	\sin[\theta(\tau)] = \frac{1}{2}\frac{\eta}{\tau} +
	\mathcal{O}\left(\frac{\eta^3}{\tau^3}\right).
	\label{eq:sineta2}
\end{equation}

We find that the leading order correction is given by 
\begin{equation}
	\begin{aligned}
		P_{1\to 1}^{(1)} &= \frac{\eta}{\sqrt{2}} \sqrt{P_{\mm{LZ}}} \sqrt{1 - P_{\mm{LZ}}} \times \\
		&\phantom{={}} 
		\left\{ -2\frac{\cos[\xi (\tau) - \xi (\tau_\ui)]}{\tau_\up} \left[
		P_{\mm{LZ}} (\sin[ \chi_1 (\tau, \tau_\ui) ] - \cos[ \chi_1 (\tau, \tau_\ui)] )
		\right. \right.\\
		&\phantom{= -2\frac{\cos[\xi (\tau) - \xi (\tau_\ui)]}{\tau_\up} [ {}} 
		\left . +
		\left( 1 - P_{\mm{LZ}}\right)( \cos[\chi_2 (\tau, \tau_\up, \tau_\ui)] +\sin[\chi_2
		(\tau, \tau_\up, \tau_\ui)]) \right] \\
		&\phantom{= [ {}}
		+ 2\frac{\cos[\xi (\tau) - \xi (\tau_\up)]}{\tau} \left[
		P_{\mm{LZ}} (\sin[ \chi_3 (\tau, \tau_\up) ] + \cos[ \chi_3 (\tau, \tau_\up)] ) 
		\right. \\
		&\phantom{= -2\frac{\cos[\xi (\tau) - \xi (\tau_\ui)]}{\tau_\up} [ {}}
		\left. + 
		\left( 1 - P_{\mm{LZ}}\right)(\cos[\chi_4 (\tau, \tau_\up)] - \sin[\chi_4
		(\tau, \tau_\up)]) \right] \\
		&\phantom{= [ {}}
		- 2\frac{\sin[\xi (\tau_\ui) - \xi (\tau_\up)]}{\tau_\ui} \left[
		P_{\mm{LZ}} 2 \cos[\xi (\tau_\ui) - \xi (\tau_\up)] (\cos[\chi_5(\tau_\up)]+
		\sin[\chi_5(\tau_\up)]) \right. \\
		&\phantom{= -2\frac{\cos[\xi (\tau) - \xi (\tau_\ui)]}{\tau_\up} [ {}}
		\left.
		-(\cos[\chi_6 (\tau_\ui, \tau_\up)] + \sin[\chi_6 (\tau_\ui, \tau_\up)]) \right] 
		\bigg\},
	\end{aligned}
	\label{eq:P111}
\end{equation}
where we have defined 
\begin{equation}
	\begin{aligned}
		\chi_1 (\tau, \tau_\ui) &= \xi (\tau) + \xi (\tau_\ui) 
		- \mm{arg}\left[\Gamma\left(1+i \frac{\eta^2}{4}\right)\right], \\
		\chi_2 (\tau, \tau_\up, \tau_\ui) &= \xi (\tau) + \xi (\tau_\ui) + 4 \xi (\tau_\up)
		- 3 \mm{arg}\left[\Gamma\left(1+i \frac{\eta^2}{4}\right)\right], \\
		\chi_3 (\tau, \tau_\up) &= \xi (\tau) - 3 \xi (\tau_\up) + 
		 \mm{arg}\left[\Gamma\left(1+i \frac{\eta^2}{4}\right)\right], \\
		 \chi_4 (\tau, \tau_\up) &= \xi (\tau) + \xi (\tau_\up) 
		 - \mm{arg}\left[\Gamma\left(1+i \frac{\eta^2}{4}\right)\right], \\
		 \chi_5 (\tau_\up) &= 2  \xi (\tau_\up) -
		 \mm{arg}\left[\Gamma\left(1+i\frac{\eta^2}{4}\right)\right], \\
		 \chi_6 (\tau_\ui, \tau_\up) &= \xi (\tau_\ui) + \xi (\tau_\up) 
		 - \mm{arg}\left[\Gamma\left(1+i\frac{\eta^2}{4}\right)\right],
	\end{aligned}
	\label{eq:phases1storder}
\end{equation}
and $\xi (\tau)$ is defined in Eq.~\eqref{eq:defasymptoticsDnz}. 

If we define (see Eq.~\eqref{eq:asymptrprob} in the main text)
\begin{equation}
	P_{1\to 1}^{(0)} = 1 - 4 P_{\mm{LZ}} \left(1-P_{\mm{LZ}}\right) \cos^2 \left[\chi_{\mm{dp}}
	(\tau_\up) \right]=P_{1\to 1}^\mathrm{inf},
	\label{eq:P110}
\end{equation}
then the return probability to leading order in $\eta/\tau, \eta/\tau_\ui, \eta/\tau_\up$ is given
by 
\begin{equation}
	P_{1 \to 1} =  P_{1\to 1}^{(0)} +  P_{1\to 1}^{(1)} +
	\mathcal{O}\left(\frac{\eta^2}{\tau^2}, \frac{\eta^2}{\tau_\up^2}, \frac{\eta^2}{\tau_\ui^2}\right) 
	\label{eq:returnProbwith1st}
\end{equation}

\section{Voltage ramps}
\label{ramp}
\setcounter{figure}{0}
The experimentally applied triangular voltage ramps are created numerically and fed to an Arbitrary Function Generator (AFG). A schematic of the applied ramps is depicted in Fig.\,\ref{Fig_App_1}.
\begin{figure}[!htb]
\includegraphics{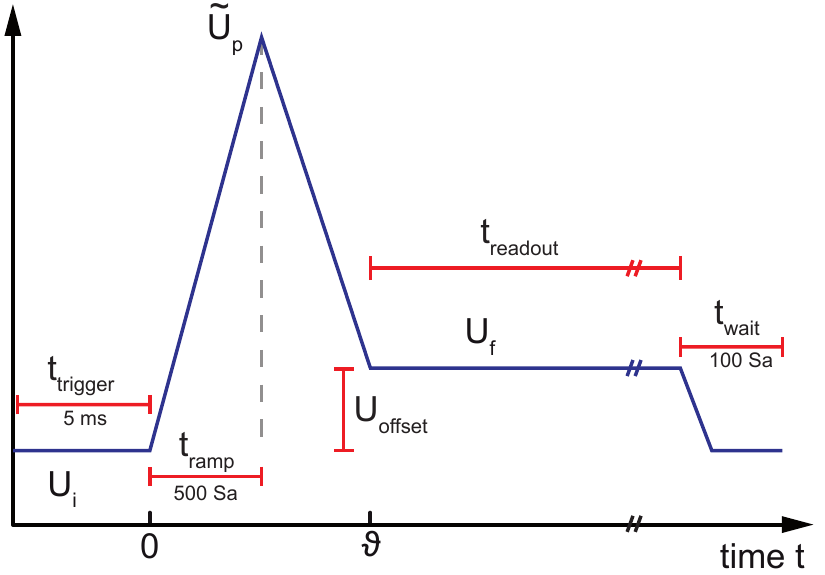}
\caption{\label{Fig_App_1} Illustration of the applied voltage ramps. The time axis is truncated as a guide to the eye since the read-out time $t_\mathrm{readout}$ is much longer than the ramp time $t_\mathrm{ramp}$.}
\end{figure}
The ramps consist of a total of 500,000\,samples (500\,kSa) divided into four basic regions. The first region is a $t_\mathrm{trigger}=5$\,ms long window in which a trigger command is sent from to the AFG to the spectrum analyzer to start the measurement, during which the additional ramp voltage is kept at zero and hence the absolute voltage is at base level $U_\mathrm{i}$ of the initialization voltage. The triangular voltage ramp itself (region two) consists in total of 1,000\,samples (1\,kSa), with 500\,Sa per ramp flank. The sweep voltage is ramped up from zero to the peak voltage $U_\mathrm{p}$ with sample rate
\begin{equation}
\label{eq:samplerate}
\mathrm{Samplerate}=\frac{500\,\mathrm{Sa}}{U_\mathrm{p} \times 1/\beta},
\end{equation}
from which we deduce the inverse sweep rate $1/\beta$. The corresponding ramp time is hence given by
\begin{equation}
\label{eq:ramptime}
t_\mathrm{ramp}=\frac{500\,\mathrm{Sa}}{\mathrm{Samplerate}}=U_\mathrm{p} \times 1/\beta.
\end{equation}
The right hand side flank of the triangular voltage ramp decreases the absolute voltage from $\widetilde{U}_\mathrm{p}$ to the read-out voltage $U_\mathrm{f}$, which is off-set from the initialization voltage $U_\mathrm{i}$ by $U_\mathrm{offset}=0.5$\,V. As described in the main text, the exponential decay of the returning excitation has to be measured at a different read-out frequency since the resonant sinusoidal drive tone at fixed frequency $\omega_1(U_\mathrm{i})/2\pi$ cannot be turned off during the voltage ramp. Hence, the above introduced voltage off-set is employed. It is important to note that the voltage off-set has to be adjusted in such a way, that the mechanical resonance at the read-out voltage $\omega_1(U_\mathrm{f})/2\pi$ is not excited by the resonant drive tone at $\omega_1(U_\mathrm{i})/2\pi$. The exponential decay of the mechanical resonance after the triangular voltage ramp is measured in region three using a spectrum analyzer in a timespan of
\begin{equation}
\label{eq:readout}
\begin{aligned}
t_\mathrm{readout}&= t_\mathrm{ramp} \times \frac{\mathrm{Total\,Samples}}{500\,\mathrm{Sa}}-2 \times t_\mathrm{ramp} \\
&= U_\mathrm{p} \times 1/\beta \left( \frac{500\,\mathrm{kSa}}{500\,\mathrm{Sa}}-2 \right).
\end{aligned}
\end{equation}
After the measurement, the absolute voltage is ramped back from $U_\mathrm{f}$ to the initialization voltage $U_\mathrm{i}$ (region four) by decreasing the sweep voltage from $U_\mathrm{offset}$ to zero, which takes 100\,samples of the total sample number of 500\,kSa.
\section{Moving average}
\label{ma}
A moving average, also referred to as sliding average, is a statistical tool for the smoothing of datasets. Consider a dataset of $N$ elements. Then, a moving average of $M$ points creates $N-M$ subsets of elements, which are averaged individually. For each element $n\geq M$ of dataset $N$, the moving average yields the mean of the subset which consists of element $n$ and the preceding $M-1$ elements in the dataset:
\begin{equation}
\label{eq:ma}
\bar{p}_\mathrm{n}=\frac{1}{M}\sum\limits_{i=0}^{M-1}p_\mathrm{n-i}
\end{equation} 
A moving average over 10 points is applied to the experimental data to extract the measurement visibility. The effect of a moving average of 10 points on the experimental data is exemplarily depicted in Fig.\,\ref{Fig_5}\,a.
\end{appendix}

\begin{acknowledgments}
Financial support by the Deutsche Forschungsgemeinschaft via the collaborative research center SFB
767 and via project Ko\,416-18 is gratefully acknowledged. H.\,R. acknowledges funding from the
Swiss SNF.
\end{acknowledgments}

%\bibliography{finite_times_v6}
%

\end{document}